\documentclass[10pt,a4paper,onecolumn,twoside]{article}
\usepackage[top=2cm,bottom=2cm,left=2cm,right=2cm]{geometry}

\usepackage[utf8]{inputenc}
\usepackage[T1]{fontenc}

\usepackage{fourier}

\usepackage{amsmath}
\allowdisplaybreaks
\usepackage{booktabs}
\usepackage{graphicx}

\usepackage{amsfonts}
\usepackage{amssymb}
\usepackage{stmaryrd}
\usepackage{mathrsfs}
\usepackage{bm}

\usepackage{amsthm}
\theoremstyle{remark}

\theoremstyle{theorem}

\usepackage{framed}
\usepackage[hidelinks]{hyperref}

\title{A hyperbolic framework for shear sound beams in nonlinear solids}
\author{~\\ Harold Berjamin\textsuperscript{a}, Michel Destrade\textsuperscript{a}\\ \\
	\small \emph{\textsuperscript{a}School of Mathematics, Statistics and Applied Mathematics,}\\ \small \emph{ NUI Galway, University Road, Galway, Republic of Ireland}}
\date{}

\usepackage[numbers,sort&compress]{natbib}
\begin{document}
	
\pagestyle{headings}
\thispagestyle{empty}
\maketitle

\begin{abstract}
\noindent
In soft elastic solids, directional shear waves are in general governed by coupled nonlinear KZK-type equations for the two transverse velocity components, when both quadratic nonlinearity and cubic nonlinearity are taken into account.
Here we consider spatially two-dimensional wave fields. We propose a change of variables to transform the equations into a quasi-linear first-order system of partial differential equations. Its numerical resolution is then tackled by using a path-conservative MUSCL-Osher finite volume scheme, which is well-suited to the computation of shock waves. We validate the method against analytical solutions (Green's function, plane waves). 
The results highlight the generation of odd harmonics and of second-order harmonics in a Gaussian shear-wave beam. \\


\end{abstract}



\section{Introduction}\label{sec:Intro}

Understanding the propagation and diffraction of sound emitted by a directional source has been an important concern of the nonlinear acoustics community since the late 1960s \cite{rudenko10}. For this purpose, the celebrated Khokhlov--Zabolotskaya--Kuznetsov (KZK) equation was derived from the equations of fluid dynamics by introducing an appropriate scaling. Valid in the paraxial region of a directive acoustic source (e.g., a transducer), this nonlinear parabolic equation describes how sound beams spread with increasing propagation distance, incorporating harmonic generation and attenuation due to nonlinearity and dissipation effects. The same approach was used for the equations of nonlinear Lagrangian elastodynamics, leading to similar partial differential equations (see the review by Norris \cite{norris98}).

In soft incompressible solids, the experimental observation of nonlinear shear waves has been reported in the literature \cite{catheline03,renier08}. Along with these observations, the generation of mainly odd harmonics and shocks has also been reported. To explain these features, Zabolotskaya et al. \cite{zabo04} showed that plane shear waves with a single transverse displacement component are governed by a Burgers-like equation when cubic nonlinearity is taken into account and quadratic nonlinearity is ignored. 
In that case, the shear waves are linearly polarized and the motion is purely anti-plane.

However, incompressible solids with cubic nonlinearity only, and no quadratic nonlinearity, are modelled  by a very special constitutive law \cite{destrade11}, not representative of real-world materials. This limitation is resolved by considering transverse shear waves with an arbitrary polarization. 
Then, two coupled KZK-type equations are obtained \cite{wochner08}. If the corresponding wave fields have variations in the transverse direction (that is, if the plane-wave assumption is relaxed), then the governing equations of motion include both quadratic and cubic nonlinearity. Related works show that the second harmonic can be generated in this configuration \cite{destrade19}, which is in agreement with more recent measurements \cite{espindola15}.

No analytical solution is known for this system of coupled nonlinear partial differential equations, and most of the above-mentioned studies rely on quasi-analytical approaches. 
Hence, Zabolotskaya et al.~\cite{zabo04} estimate the generation of harmonics by using a space-dependent harmonic expansion, and by performing harmonic balance. Wochner et al.~\cite{wochner08} use a similar harmonic expansion along with Green's function expansions. 
Finally, Destrade et al.~\cite{destrade19} implement a perturbation method based on a small amplitude parameter.

While analytical results are of great interest, computational approaches may be more versatile.
The numerical resolution of KZK-type equations was addressed by Hamilton et al.~\cite{hamilton85}, by matching near-field and far-field Fourier series expansions (Bergen code). Other frequency-domain approaches \cite{christopher91,khokhlova01} were then followed by time-domain methods, in particular by making use of operator splitting \cite{lee95}. 
Pinton and Trahey~\cite{pinton08} combined operator splitting with shock-capturing Godunov-type methods to provide accurate shock-wave solutions. To the present authors' knowledge, no method has yet been successful in solving the quadratic-cubic nonlinear system \cite{wochner08} describing directional shear-wave motion in soft elastic solids.

In this article, we consider spatially two-dimensional wave fields. After a brief presentation of the governing equations, we introduce a change of unknowns and of dependent variables that transforms the system at hand into a quasi-linear system of first-order partial differential equations (Section~\ref{sec:Prob}). In particular, the physical time variable is used instead of the retarded time. Since the differential system so-obtained is non-conservative, particular care is required when computing shock-wave solutions. Indeed, a naive upwind scheme would lead to inaccurate wave speeds \cite{leveque02}. In this study, we implement a finite volume method based on the path-conservative Osher Riemann solver and on MUSCL reconstruction \cite{dumbser11, toro09} (Section~\ref{sec:Numer}). Although it does not involve operator splitting, the scheme accounts naturally for nonlinearity, coupled motion and beam diffraction. 
We validate the method by using dedicated analytical solutions which are summarized in the \ref{app:Fund} (Green's function, plane waves). Numerical simulations of directional wave beams illustrate the generation of odd and second-order harmonics (Section~\ref{sec:Harmonic}), as predicted by Destrade et al. \cite{destrade19}.

One benefit of the first-order formulation introduced in this study is the potential to use advanced computational methods, including high-order adaptive schemes based on ADER or WENO approaches (see Refs.~\cite{castro06,reinarz20} and references therein). Moreover, such differential systems of hydrodynamic type have been studied extensively, and dedicated integrability criteria are known \cite{ferapontov06}. Prospective applications encompass the study of traumatic brain injury \cite{espindola15}, as well as related medical imaging techniques.


\section{Problem statement}\label{sec:Prob}


\subsection{Governing equations}

We introduce the deformation gradient tensor  $\bm{F} = {\partial \bm{x}}/{\partial \bm{X}}$,
where $\bm x$ represents the position of a particle in the deformed configuration, and $\bm X$ represents its position in the undeformed configuration \cite{ogden84,norris98,holzapfel00}. The Lagrangian specification of motion is used throughout the present document, so that spatial differential operators are always computed with respect to $\bm{X}$. The components $\bm{X} = (X,Y,Z)$ of the position are expressed with respect to an orthonormal basis $(\bm{e}_1,\bm{e}_2,\bm{e}_3)$ of the Euclidean space, and a Cartesian coordinate system is chosen. 
Introducing the displacement field $\bm{u} = \bm{x}-\bm{X} = (u_1,u_2,u_3)$, we write the deformation gradient as $\bm{F} = \bm{I} + \text{grad}\, \bm{u}$, where $\bm{I}$ is the identity tensor. Consequently, we also have
\begin{equation}
	\partial_t \bm{F} = \text{grad}\, \bm{v} \, ,
	\label{Kinematics}
\end{equation}
where $\bm{v} = \partial_t \bm{u}$ is the particle velocity.

In this paper we consider \emph{incompressible} hyperelastic materials, for which the constraint of no volume dilatation
\begin{equation}
 \det\bm{F}  \equiv 1
	\label{Incomp}
\end{equation}
is prescribed at all times, so that the mass density $\rho$ is constant.
The deformation is also governed by the equation of motion \cite{ogden84,norris98,holzapfel00}
\begin{equation}
	\rho\, \partial_t \bm{v} = \text{div}\, \bm{P} + \bm{f} ,
	\label{EqMot}
\end{equation}
where $\bm{f}$ is the density of body force per unit volume. The dependence of the first Piola--Kirchhoff stress tensor $\bm P$ with $\bm F$ is specified by the constitutive law.

For incompressible solids, the constitutive law may be expressed as $\bm{P} = -p\bm{F}^{-\top}\! + {\partial W}/{\partial \bm F}$,
where $p$ is a Lagrange multiplier due to incompressibility and $W$ is the strain energy density.
For instance, the strain energy of homogeneous and isotropic incompressible solids may be expanded as \cite{wochner08,destrade19}
\begin{equation}
	W = \mu I_2 + \tfrac13 A I_3 + D {I_2}^2
	\label{WZabo}
\end{equation}
in terms of the invariants $I_k = \text{tr}\, \bm{E}^k$ of the Green--Lagrange strain tensor
\begin{equation}
	\bm{E} = \tfrac12 \big[\text{grad}\, \bm{u} + \text{grad}^\top\! \bm{u} + (\text{grad}^\top\! \bm{u})( \text{grad}\, \bm{u})\big] \, .
\end{equation}
This finite-strain tensor is linked to the right Cauchy--Green deformation tensor $\bm{C} = \bm{F}^\top \bm{F}$ through the relation $\bm{E} = \frac12(\bm{C} - \bm{I})$.
The material parameters of \eqref{WZabo} are the initial shear modulus $\mu$ (second Lam\'e coefficient) and the higher-order elastic constants $A$, $D$ (Landau constants of quadratic and cubic nonlinear elasticity, respectively).

The constitutive law may be rewritten as $\bm{P} = \bm{F} \bm{S}$, where $\bm{S} = -p\bm{C}^{-1}\! +  \partial W/\partial \bm E$ is the second Piola--Kirchhoff stress tensor. 
Note that when computing the tensor derivative $\partial W/\partial \bm E$, we must keep in mind that the incompressibility constraint introduces a dependence of one invariant $I_k$ with respect to the two others (see \cite{destrade2010onset} and  \ref{app:Invar}).  In the Appendix, we derive the expression of the Cauchy stress tensor $\bm \sigma$ as
\begin{equation}
	\bm{\sigma} =  -p\bm{I} + 2\left(\frac{\partial W}{\partial \textit{I}_{\bm C}} + \textit{I}_{\bm C} \frac{\partial W}{\partial \textit{II}_{\bm C}}\right) \bm{B} - 2\frac{\partial W}{\partial \textit{II}_{\bm C}}\bm{B}^2
	\label{ConstSigB}
\end{equation}
in terms of $\bm{B} = \bm{F} \bm{F}^\top\!$, the left Cauchy--Green deformation tensor, 
from which the expression of $\bm{P} = \bm{\sigma} \bm{F}^{-\top}\!$ is deduced. The coefficients in Eq.~\eqref{ConstSigB} are detailed in the Appendix (Eq.~\eqref{ConstSCCoeffs}). 
Using the Cayley--Hamilton theorem, the previous stress-strain relationship may be written in terms of $\bm{I}$, $\bm{B}$ and $\bm{B}^{-1}$ up to a redefinition of the arbitrary Lagrange multiplier \cite{destrade19}. 
However, we keep the present form to avoid the computation of inverse matrices when working out the equations of motion later on.


\subsection{Scaling the equations of motion}

\begin{figure}[h!]
	\centering
	\includegraphics{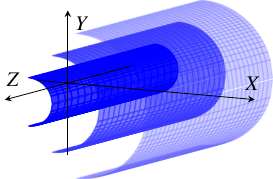}
	
	\caption{Typical directional wave fields of this study; Sketch of equal-phase surfaces issued from a line source located on the $Z$-axis. \label{fig:Sym}}
\end{figure}

Similarly to Wochner et al.~\cite{wochner08}, we introduce the following scaling
\begin{equation}
	{\renewcommand{\arraystretch}{2}
	\begin{array}{l}
		u_1 = \epsilon^2 U_1, \quad
		u_2 = \epsilon U_2, \quad
		u_3 = \epsilon U_3, \quad
		p = \epsilon^2 \tilde p , \quad
		\bm{f} = \epsilon^3 \tilde{\bm f} , \\
		\tilde{X} = \epsilon^2 X, \quad
		\tilde{Y} = \epsilon Y , \quad
		\tilde t = t - X/c ,
	\end{array}}
	\label{EpsilonPert}
\end{equation}
where $\epsilon$ is a small dimensionless parameter and $c = \sqrt{\mu/\rho}$ is the shear wave speed of linear elasticity. No scaling is assumed for the $Z$ coordinate, as the field variables are assumed invariant with respect to $Z$ (see Fig.~\ref{fig:Sym}). Here, the fields $\bm{u}$, $p$, $\bm f$ are functions of the coordinates ${\bm X}, t$, while the new variables $\bm{U}$, $\tilde p$, $\tilde{\bm f}$ depend on $\tilde{\bm X}, \tilde t$. Note that the present paraxial approximation for directive sources differs significantly from the geometric acoustics/optics approximation (ray theory) \cite{nunez20}, even though formal similarities may be found.

Following the transformation rules \eqref{EpsilonPert}, we rewrite the variables $\bm F$, $\bm v$ and the equations of motion \eqref{Kinematics}-\eqref{EqMot}  as
\begin{equation}
	\begin{aligned}
	\bm{F}  &=  \bm{I} + \partial_{\bm{U}} \bm{u} \cdot \big(\widetilde{\text{grad}}\, \bm{U} \cdot \partial_{\bm X} \tilde{\bm{X}} + \partial_{\tilde t} \bm{U} \otimes \partial_{\bm X} \tilde{t} \big) &\qquad \partial_{\tilde t}\bm{F} &= \widetilde{\text{grad}}\, \bm{v} \cdot \partial_{\bm X} \tilde{\bm{X}} + \partial_{\tilde t} \bm{v} \otimes \partial_{\bm X} \tilde{t} \\
	\bm{v} &= \partial_{\bm{U}} \bm{u} \cdot \partial_{\tilde t} \bm{U} &\qquad \rho\, \partial_{\tilde t}\bm{v} &= \widetilde{\text{grad}}\, \bm{P} : (\partial_{\bm X} \tilde{\bm{X}})^\top \! + \partial_{\tilde t} \bm{P} \cdot \partial_{\bm X} \tilde{t} + \bm{f}
	\end{aligned}
	\label{EpsilonTransf}
\end{equation}
in terms of the new coordinates, where the Jacobian matrices have the following components:
\begin{equation}
	\partial_{\bm{U}} \bm{u} =
	\begin{bmatrix}
	\epsilon^2 & 0 & 0 \\
	0 & \epsilon & 0 \\
	0 & 0 & \epsilon 
	\end{bmatrix} ,
	\qquad
	\partial_{\bm X} \tilde{\bm{X}} =
	\begin{bmatrix}
	\epsilon^2 & 0 & 0 \\
	0 & \epsilon & 0 \\
	0 & 0 & 1
	\end{bmatrix} ,
	\qquad\text{and}\qquad
	\partial_{\bm X} \tilde{t} =
	\begin{bmatrix}
	-1/c \\
	0 \\
	0
	\end{bmatrix} .
	\label{EpsilonJacobi}
	\end{equation}
	In terms of the displacement field $\bm U$, we have
	\begin{equation}
	\bm{F} = \bm{I} +
	\begin{bmatrix}
	\epsilon^2 (\epsilon^2 U_{1,\tilde 1} - \tfrac1{c} U_{1, \tilde t}) & \epsilon^3 U_{1,\tilde 2} & 0 \\[4pt]
	\epsilon   (\epsilon^2 U_{2,\tilde 1} - \tfrac1{c} U_{2, \tilde t}) & \epsilon^2 U_{2,\tilde 2} & 0 \\[4pt]
	\epsilon   (\epsilon^2 U_{3,\tilde 1} - \tfrac1{c} U_{3, \tilde t}) & \epsilon^2 U_{3,\tilde 2} & 0
	\end{bmatrix}
	\qquad\text{and}\qquad
	\begin{aligned}
	\rho \epsilon^2\, {U}_{1,\tilde t \tilde t} &= \epsilon^2 {P}_{11,\tilde 1} + \epsilon {P}_{12,\tilde 2} - \tfrac1{c} {P}_{11, \tilde t} + \epsilon^3 \tilde f_1 \\
	\rho \epsilon\, {U}_{2, \tilde t \tilde t} &= \epsilon^2 {P}_{21,\tilde 1} + \epsilon {P}_{22,\tilde 2} - \tfrac1{c} {P}_{21,\tilde t} + \epsilon^3 \tilde f_2 \\
	\rho \epsilon\, {U}_{3, \tilde t \tilde t} &= \epsilon^2 {P}_{31,\tilde 1} + \epsilon {P}_{32,\tilde 2} - \tfrac1{c} {P}_{31, \tilde t} + \epsilon^3 \tilde f_3 \, ,
	\end{aligned}
\end{equation}
where the components of $\bm P$ are deduced from the scaled components of the deformation gradient tensor $\bm F$. Here, partial differentiation is specified using subscript notation (after the commas). Integer subscripts $\tilde 1$-$\tilde 3$ denote partial differentiation with respect to the components of the position vector $\tilde{\bm X}$, while the subscript $\tilde t$ denotes differentiation in time.
At leading (quadratic) order in $\epsilon$, the incompressibility constraint \eqref{Incomp} amounts to the substitution $U_{1,\tilde t} = c U_{2,\tilde 2}$ in the above equations. 

By transforming back to the original spatial coordinates $\bm X$ and by making appropriate substitutions, the same equations as Eqs. (10)-(11) of \cite{wochner08} are obtained at cubic order as
\begin{equation}
	\begin{aligned}
	u_{2, 1 \tilde t} &= \frac{c}{2} (\alpha^2 u_{2, 22} + f_2/\mu)
	+ \frac{\beta_2}{2 c} \big(u_{3, \tilde t \tilde t} u_{3, 2} - u_{3, 2\tilde t}u_{3,\tilde t}\big)
	+ \frac{\beta_3}{3 c^3} \big(u_{2,\tilde t}({u_{2,\tilde t}}^2 + {u_{3,\tilde t}}^2)\big)_{,\tilde t} \\
	u_{3, 1 \tilde t} &= \frac{c}{2} (\alpha^2 u_{3, 22} + f_3/\mu)
	+ \frac{\beta_2}{2 c} \big(u_{2,\tilde t\tilde t}u_{3,2} - u_{2,2\tilde t}u_{3,\tilde t}\big) + \frac{\beta_2}{c} \big(u_{2,\tilde t}u_{3,2\tilde t} - u_{2,2}u_{3,\tilde t\tilde t}\big)  + \frac{\beta_3}{3 c^3} \big(u_{3,\tilde t}({u_{2,\tilde t}}^2 + {u_{3,\tilde t}}^2)\big)_{,\tilde t}
	\end{aligned}
	\label{SystWochner}
\end{equation}
but with additional spatial symmetries due to invariance along the $Z$-coordinate. Here, the displacement $\bm{u}$ depends on the coordinates ${\bm X}, \tilde t$, which is standard but slightly abusive notation compared to the initial definitions \eqref{EpsilonPert}. At leading (quadratic) order, the  equation giving the Lagrange multiplier in unbounded domain reads $p = \rho \beta_2 \big({u_{2,\tilde t}}^2 + {u_{3,\tilde t}}^2\big)$. In Eq.~\eqref{SystWochner},  we introduced the  quadratic terms  coefficient $\beta_2$ and the cubic terms  coefficient $\beta_3$ \cite{wochner08,destrade19}
\begin{equation}
	\beta_2 = 1 + \frac{A}{4\mu},
	\qquad
	\beta_3 = \frac32 \left(1 + \frac{A/2 + D}{\mu}\right) .
	\label{Beta}
\end{equation}
For later use, we  introduced a parameter $\alpha\in\lbrace 0, 1\rbrace$ in Eq.~\eqref{SystWochner} that gives the possibility to discard the diffraction term. Note that if the configuration is invariant along the transverse $Y$-axis, then the diffraction term vanishes as well as the quadratic term \cite{zabo04,renier08}.


\subsection{First-order recast}\label{sec:FirstOrder}

Let us introduce the displacement's partial derivatives $v_i = u_{i,\tilde t}$, $\theta_i = u_{i, 1}$, $\varepsilon = u_{2, 2}$ and $\vartheta = u_{3, 2}$ with $i= 2,3$. Thus, the system \eqref{SystWochner} is rewritten as
\begin{equation}
	\begin{aligned}
	\theta_{2,\tilde t} &= \frac{c}{2} (\alpha^2\varepsilon_{, 2} + f_2/\mu)
	+ \frac{\beta_2}{2 c} \big(v_{3,\tilde t} \vartheta - \vartheta_{,\tilde t}v_{3}\big)
	+ \frac{\beta_3}{3 c^3} \big(v_2({v_2}^2 + {v_3}^2)\big)_{,\tilde t} \\
	\theta_{3,\tilde t} &= \frac{c}{2} (\alpha^2\vartheta_{,2} + f_3/\mu)
	+ \frac{\beta_2}{2 c} \big(v_{2,\tilde t} \vartheta - \varepsilon_{,\tilde t}v_{3} + 2(v_2\vartheta_{,\tilde t} - \varepsilon v_{3,\tilde t})\big)
	+ \frac{\beta_3}{3 c^3} \big(v_3({v_2}^2 + {v_3}^2)\big)_{,\tilde t} \, .
	\end{aligned}
	\label{SystWochnerRecast}
\end{equation}
By making use of the equality of mixed partial derivatives, four kinematic relationships between the strains $u_{i,j}$ and the velocities $u_{i,\tilde t}$ are derived. The equations form a first-order PDE system $\mathbf{a} \mathbf{q}_{,1} + \mathbf{b} \mathbf{q}_{,2} + \mathbf{c}(\mathbf{q}) \mathbf{q}_{,\tilde t} = \mathbf{s}$ in terms of the vector of unknowns $\mathbf{q} = (v_2, \theta_2, \varepsilon, v_3, \theta_3, \vartheta)^\top\!$. The matrix $\mathbf{c}(\mathbf{q}) = \mathbf{c}_\text{L} + \mathbf{c}_\text{NL}(\mathbf{q})$ is decomposed as the sum of a constant part $\mathbf{c}_\text{L}$ and of a non-constant part $\mathbf{c}_\text{NL}(\mathbf{q})$, which vanishes if the parameters of nonlinearity $\beta_2$, $\beta_3$ are zero. The corresponding matrices are detailed in the \ref{app:Matrices}. 
Note that the matrix $\mathbf{c}_\text{NL}(\mathbf{q})$ does not depend on $\theta_2$, $\theta_3$, and that it vanishes if $\mathbf{q} \to \mathbf{0}$.

In Eqs.~\eqref{SystWochner}-\eqref{SystWochnerRecast}, the unknowns $\bm u$, ${\bf q}$ are functions of $\bm X$ and $\tilde t$ (explicit dependence has been dropped for sake of conciseness). Now, the transformation from the retarted time $\tilde t$ to the real time $t$ is carried out. Thus, we introduce the displacement field ${\bm u}(\bm{X}, \tilde t) = {\bm \eta}({\bm X}, t)$. Using differentiation rules, one shows that $\bf q$ satisifes $v_i = \eta_{i,t}$ and $\theta_i = \eta_{i, 1} + v_i/c$, while the differential definitions of $\varepsilon$, $\vartheta$ are the same whether $\bm u$ or $\bm \eta$ is used. Next, we introduce the vector $\mathbf{p} = \mathbf{T}^{-1} \mathbf{q} = (v_2, \gamma_2, \varepsilon, v_3, \gamma_3, \vartheta)^\top\!$ such that $\gamma_i = \eta_{i,1}$. The transformation matrix $\mathbf{T}$ is defined in such a way that the variables $\mathbf{q} = \mathbf{T}\, \mathbf{p}$ are modified according to $\gamma_i = \theta_i - v_i/c$.

The first-order PDE system reads $\mathbf{a}' \mathbf{p}_{,1} + \mathbf{b}' \mathbf{p}_{,2} + \mathbf{c}'(\mathbf{q}) \mathbf{p}_{, t} = \mathbf{s}$ in terms of the original time variable $t$, where the prime denotes right-multiplication by $\mathbf{T}$ and where $\mathbf{c}'(\mathbf{q}) = \mathbf{c}'_\text{L} + \tfrac1{c}\mathbf{a}' + \mathbf{c}'_\text{NL}(\mathbf{q})$. For later use, we compute the determinant $\Delta$ of the matrix $\mathbf{c}'(\mathbf{q})$, and the expression
\begin{equation}
	c^2 \Delta = 1 + \beta_2 \varepsilon - \tfrac43 {\beta_3} \tfrac{{v_2}^2 + {v_3}^2}{c^2} - \tfrac1{4} {\beta_2}^2 \vartheta^2 - \tfrac13 \beta_2 \beta_3 \tfrac{({v_2}^2 + {v_3}^2) \varepsilon + 2 v_2 (v_2 \varepsilon + v_3 \vartheta)}{c^2}  + \tfrac13 {\beta_3}^2 \big( \tfrac{{v_2}^2 + {v_3}^2}{c^2} \big)^2 
	\label{DeterminantC}
\end{equation}
is obtained. In particular, we see that this expression is nonzero even if $\beta_2$ and $\beta_3$ are both equal to zero, or if $(v_2, \varepsilon, v_3, \vartheta)$ is sufficiently close to zero. As long as the matrix $\mathbf{c}'(\mathbf{q})$ is not singular, we can left-multiply our system by $\mathbf{c}'(\mathbf{q})^{-1}$ to rewrite the equations of motion as a \emph{quasi-linear first-order system of balance laws},
\begin{equation}
	\mathbf{p}_{, t} + \mathbf{A}(\mathbf{p})\, \mathbf{p}_{,1} + \mathbf{B}(\mathbf{p})\, \mathbf{p}_{,2}  = \mathbf{S}(\mathbf{p})
	\qquad\text{with}\qquad
	\begin{aligned}
	\mathbf{A}(\mathbf{p}) &= \mathbf{c}'(\mathbf{T}\, \mathbf{p})^{-1} \mathbf{a}' , \\
	\mathbf{B}(\mathbf{p}) &= \mathbf{c}'(\mathbf{T}\, \mathbf{p})^{-1} \mathbf{b}' , \\
	\mathbf{S}(\mathbf{p}) &= \mathbf{c}'(\mathbf{T}\, \mathbf{p})^{-1} \mathbf{s} \, .
	\end{aligned}
	\label{QLSystemCons}
\end{equation}
We give the expressions of the above matrices and vectors in  \ref{app:Matrices}.


\paragraph{Hyperbolicity}

Let us assume that $\beta_2$ and $\beta_3$ both equal zero, so that the system \eqref{SystWochner} becomes linear. In this case, the displacement fields $u_2$ and $u_3$ decouple. Moreover, they satisfy the same linear partial differential equation $u_{, 1 \tilde t} = \tfrac{c}2 (\alpha^2 u_{, 22} + f/\mu)$. According to the theory of second-order differential equations in three independent variables (see e.g. \cite{courant}, Sec. III.3.1), this equation is hyperbolic. If the real time variable $t$ is used instead of the retarded time $\tilde t$, then the previous equation may be rewritten as $(u_{, t} + c u_{, 1})_{,t} - \tfrac{1}2 \alpha^2 c^2 u_{, 22} = g$ where $g = \tfrac{1}2 f/\rho$. 
We then recognize a modified wave equation in $(Y,t)$ coordinates, where the classical term $u_{,tt}$ has been replaced by the time derivative of a transport term along $X$. Analytical solutions to this equation are detailed in \ref{app:Fund}.

Now we study the system of balance laws \eqref{QLSystemCons} where the coefficients $\beta_2$, $\beta_3$ are set to zero. The spectral properties of the matrix $\mathbf{A}$ show that the characteristic speed along the $X$-axis is $+c$. Similarly, the spectral properties of $\mathbf{B}$ yield the characteristic speeds $\pm \alpha c/\sqrt{2}$ along the $Y$-axis.
Moreover, the hyperbolicity property can be deduced from the spectrum of the linear combination $n_1\mathbf{A} + n_2\mathbf{B}$ where $\mathbf{n} = (n_1, n_2)$ is a unit vector \cite{godlewski96,leveque02}.
This property generalises to the quasi-linear system \eqref{QLSystemCons} with arbitrary $\beta_2$, $\beta_3$, as long as $(v_2, \varepsilon, v_3, \vartheta)$ stays in the vicinity of the origin.
More detailed conditions can be derived analytically in the particular case $\beta_2 = 0$, $\beta_3 > 0$, where the non-singularity of the determinant \eqref{DeterminantC} implies ${v_2}^2 + {v_3}^2 < c^2/\beta_3$. If this condition is satisfied, then the eigenvalues of $n_1\mathbf{A}(\mathbf{p}) + n_2\mathbf{B}(\mathbf{p})$ are real for any unit vector $\mathbf{n}$ in both cases $\alpha \in \lbrace 0, 1\rbrace$, so that hyperbolicity is ensured.

\bigskip
\par

We note that the equations of motion have a structure similar to Lagrangian elastodynamics. Indeed, the full system of nonlinear elasticity reads as a conservative first-order system \cite{lee13}, where the conserved variables $\mathbf{p}$ are the displacement gradients and the velocities. Nevertheless, the number of state variables is smaller than in the case of two-dimensional elastodynamics \cite{berjamin19} due to assumed deformation and symmetries. An additional remarkable feature is the \emph{one-way} nature of the wave propagation, which results from the scaling procedure. In the present directional wave beam model, a non-conservative system is naturally obtained. Specific theoretical and numerical difficulties arise with such quasi-linear systems of partial differential equations, due to the presence of the non-conservative products $\mathbf{A}(\mathbf{p})\, \mathbf{p}_{,1}$ and $\mathbf{B}(\mathbf{p})\, \mathbf{p}_{,2}$. A dedicated numerical method is presented in the next section.


\section{Numerical resolution}\label{sec:Numer}

In the examples presented later on, the physical domain is assumed unbounded. We consider a finite numerical domain for $\bm{X} = (X,Y)$ in $[-1 , 1] \times [ -1, 1] /\sqrt{2}$. It is discretized using a regular grid in space with mesh size $\Delta x$ in the $X$-direction, and $\Delta y$ in the $Y$-direction. The coordinates of the nodes are $(x_i,y_j) = (i\, \Delta x,j\, \Delta y)$, where $0\leqslant i\leqslant N_x$ and $0\leqslant j\leqslant N_y$. The total number of nodes is $(N_x+1)\times (N_y+1)$, where $N_x = 2/\Delta x$ and $N_y = \sqrt{2}/\Delta y$ denote the number of cells in each direction. A variable time step $\Delta t = t_{n+1} - t_n$ is introduced. Therefore, $\mathbf{p}(x_i,y_j, t_n)$ denotes the solution to \eqref{QLSystemCons} at the grid node $(i, j)$ and at the $n$th time step. Numerical approximations of the solution are denoted by $\mathbf{p}_{i,j}^n \simeq \mathbf{p}(x_i,y_j, t_n)$.


\subsection{Finite volume method}\label{sec:FV}

The system of balance laws
\eqref{QLSystemCons} is integrated explicitly according to the following updating formula:
\begin{equation}
	\mathbf{p}_{i,j}^{n+1} = \mathbf{p}_{i,j}^n - \frac{\Delta t}{\Delta x} \left(\mathcal{D}\mathbf{f}_{i-\frac12,j}^+ + \mathcal{D}\mathbf{f}_{i+\frac12 ,j}^- + \delta\mathbf{f}_{i,j}\right) - \frac{\Delta t}{\Delta y} \left(\mathcal{D}\mathbf{g}_{i,j-\frac12}^+ + \mathcal{D}\mathbf{g}_{i,j+\frac12}^- + \delta\mathbf{g}_{i,j}\right) + \Delta t\, \mathbf{S}_{i,j}^n ,
	\label{SchemaExplCons}
\end{equation}
where the approximation $\mathbf{S}_{i,j}^n$ of the source term $\mathbf{S}(\mathbf{p})$ is specified later on. This formula is used for the interior cells, while pseudo-absorbing boundary conditions are implemented at the boundaries of the numerical domain (Sec. 21.8.5 p.~488 of \cite{leveque02}).

As described in the next paragraph, the jumps $\mathcal{D}\mathbf{f}_{i\mp 1/2,j}^\pm$, $\mathcal{D}\mathbf{g}_{i,j\mp 1/2}^\pm$ and the corrections $\delta\mathbf{f}_{i,j}$, $\delta\mathbf{g}_{i,j}$ in \eqref{SchemaExplCons} are computed according to a path-conservative finite volume method with slope limiters \cite{castro06, toro09, dumbser11}. The method avoids spurious oscillations and is nearly second-order accurate in space and time on smooth solutions. Moreover, it does not suffer the severe limitations of the ``naive'' non-conservative upwind method regarding non-smooth solutions (see \cite{leveque02},  p.~238). Note that due to the nonlinearity of the system, shocks might form even if loadings are smooth.

The numerical flux differences are computed by applying the MUSCL--Hancock procedure componentwise~\cite{toro09}, in combination with a path-conservative Osher scheme~\cite{dumbser11}. The method consists of the following steps:
\begin{enumerate}
	\item We construct the left (`$-$') and right (`$+$') linearly extrapolated values at the cell interfaces $(x_{i + 1/2}, y_j)$ and $(x_i, y_{j + 1/2})$ as follows (Sec.~14.4.2 and 16.5 of \cite{toro09}):
	\begin{align}
			\mathbf{p}_{i+\frac12,j}^- &= \mathbf{p}_{i,j}^n + \frac12 \left(\mathbf{I} - \frac{\Delta t}{\Delta x}\mathbf{A}(\mathbf{p}_{i,j}^n) \right) \bm{\Delta}_{i} \mathbf{p}_{\bullet ,j} - \frac12 \frac{\Delta t}{\Delta y}\mathbf{B}(\mathbf{p}_{i,j}^n) \bm{\Delta}_{j} \mathbf{p}_{i ,\bullet} \, , \nonumber\\
			\mathbf{p}_{i+\frac12,j}^+ &= \mathbf{p}_{i+1,j}^n - \frac12 \left(\mathbf{I} + \frac{\Delta t}{\Delta x}\mathbf{A}(\mathbf{p}_{i+1,j}^n) \right) \bm{\Delta}_{i+1} \mathbf{p}_{\bullet ,j} - \frac12 \frac{\Delta t}{\Delta y}\mathbf{B}(\mathbf{p}_{i+1,j}^n) \bm{\Delta}_{j} \mathbf{p}_{i+1 ,\bullet} \, , \label{SlopeFinal} \\
			\mathbf{p}_{i,j+\frac12}^- &= \mathbf{p}_{i,j}^n - \frac12 \frac{\Delta t}{\Delta x}\mathbf{A}(\mathbf{p}_{i,j}^n) \bm{\Delta}_{i} \mathbf{p}_{\bullet ,j} + \frac12 \left(\mathbf{I} - \frac{\Delta t}{\Delta y}\mathbf{B}(\mathbf{p}_{i,j}^n) \right) \bm{\Delta}_{j} \mathbf{p}_{i ,\bullet} \, , \nonumber\\
			\mathbf{p}_{i,j+\frac12}^+ &= \mathbf{p}_{i,j+1}^n - \frac12 \frac{\Delta t}{\Delta x}\mathbf{A}(\mathbf{p}_{i,j+1}^n) \bm{\Delta}_{i} \mathbf{p}_{\bullet, j+1} - \frac12 \left(\mathbf{I} + \frac{\Delta t}{\Delta y}\mathbf{B}(\mathbf{p}_{i,j+1}^n) \right) \bm{\Delta}_{j+1} \mathbf{p}_{i, \bullet} \, . \nonumber
	\end{align}
	The coefficients of the matrices $\mathbf{A}(\mathbf{p})$ and $\mathbf{B}(\mathbf{p})$ are provided in  \ref{app:Matrices}.
	The vectors
	\begin{equation}
		\begin{aligned}
			\bm{\Delta}_{i} \mathbf{p}_{\bullet ,j} &= \text{MC}(\mathbf{p}_{i,j}^n - \mathbf{p}_{i-1,j}^n, \mathbf{p}_{i+1,j}^n - \mathbf{p}_{i,j}^n) \, ,  \\
			\bm{\Delta}_{j} \mathbf{p}_{i ,\bullet} &= \text{MC}(\mathbf{p}_{i,j}^n - \mathbf{p}_{i,j-1}^n, \mathbf{p}_{i,j+1}^n - \mathbf{p}_{i,j}^n) \, , 
		\end{aligned}
	\end{equation}
	are limited differences of $\mathbf{p}$ along the $X$- and $Y$-directions{\,---\,}in other words, they are limited slopes multiplied by the mesh size. Here, the monotonized central-difference limiter defined by
	\begin{equation}
		\text{MC}(\mathbf{a},\mathbf{b}) = \tfrac12 (\text{sgn}\,\mathbf{a} + \text{sgn}\,\mathbf{b}) \min \left(2\, |\mathbf{a}|, 2\, |\mathbf{b}|, \tfrac12 |\mathbf{a}+\mathbf{b}|\right)
		\label{minmod}
	\end{equation}
	is applied componentwise, where $\text{sgn}$ denotes the sign function \cite{leveque02}.

	\item The evaluation of the jumps is performed according to a path-conservative Osher scheme \cite{dumbser11}
	\begin{equation}
		\begin{aligned}
			\mathcal{D}\mathbf{f}_{i+\frac12,j}^\pm &= \left( \int_0^1 \mathbf{A}^\pm \big(s \mathbf{p}_{i+\frac12,j}^+ + (1-s) \mathbf{p}_{i+\frac12,j}^-\big) \,\text d s \right) (\mathbf{p}_{i+\frac12,j}^+ - \mathbf{p}_{i+\frac12,j}^-) \, , \\
			\mathcal{D}\mathbf{g}_{i+\frac12,j}^\pm &= \left( \int_0^1 \mathbf{B}^\pm \big(s \mathbf{p}_{i,j+\frac12}^+ + (1-s) \mathbf{p}_{i,j+\frac12}^-\big) \,\text d s \right) (\mathbf{p}_{i,j+\frac12}^+ - \mathbf{p}_{i,j+\frac12}^-) \, ,
		\end{aligned}
		\label{Osher}
	\end{equation}
	where a linear integration path is used.
	A spectral decomposition of the matrix
	\begin{equation}
		\mathbf{A}(\mathbf{p}) = \mathbf{R}(\mathbf{p})\, \mathbf{\Lambda}(\mathbf{p})\, \mathbf{R}(\mathbf{p})^{-1}
		\label{EigenA}
	\end{equation}
	is obtained numerically (as well as a spectral decomposition of $\mathbf{B}(\mathbf{p})$), e.g. by using the \texttt{eigen} function of the Julia Language \cite{julialang}. Here, $\mathbf{R}(\mathbf{p})$, $\mathbf{\Lambda}(\mathbf{p})$ denote the matrices of right eigenvectors of $\mathbf{A}(\mathbf{p})$ and corresponding eigenvalues, respectively. The matrices
	\begin{equation}
		\mathbf{A}^\pm(\mathbf{p}) = \mathbf{R}(\mathbf{p})\, \mathbf{\Lambda}^\pm(\mathbf{p})\, \mathbf{R}(\mathbf{p})^{-1}
		\label{EigenApm}
	\end{equation}
	and $\mathbf{B}^\pm(\mathbf{p})$ are obtained by taking the positive part `$+$' or the negative part `$-$' of the eigenvalues to ensure correct upwinding. The corrections
	\begin{equation}
		\begin{aligned}
		\delta\mathbf{f}_{i,j} &= \left( \int_0^1 \mathbf{A} \big(s \mathbf{p}_{i+\frac12,j}^- + (1-s) \mathbf{p}_{i-\frac12,j}^+\big) \,\text d s \right) (\mathbf{p}_{i+\frac12,j}^- - \mathbf{p}_{i-\frac12,j}^+) \, , \\
		\delta\mathbf{g}_{i,j} &= \left( \int_0^1 \mathbf{B} \big(s \mathbf{p}_{i,j+\frac12}^- + (1-s) \mathbf{p}_{i,j-\frac12}^+\big) \,\text d s \right) (\mathbf{p}_{i,j+\frac12}^- - \mathbf{p}_{i,j-\frac12}^+) \, ,
		\end{aligned}
		\label{OsherCorr}
	\end{equation}
	are designed to ensure consistency with conservative methods \cite{castro06}. Note that these important terms are not included in Ref.~\cite{dumbser11}. Similarly to \eqref{Osher}, a linear integration path was used.
	The integrals \eqref{Osher}--\eqref{OsherCorr} are computed numerically by using the three-point Gauss--Legendre quadrature rule~\cite{dumbser11}, which proved sufficient to reach second-order accuracy.
\end{enumerate}
Fromm-type methods are recovered by replacing the MC function \eqref{minmod} with the linear average $(\mathbf{a},\mathbf{b}) \mapsto \frac12(\mathbf{a}+\mathbf{b})$, while the projections $(\mathbf{a},\mathbf{b}) \mapsto \mathbf{a}$ or $\mathbf{b}$ yield either Beam--Warming or Lax--Wendroff-type methods~\cite{toro09}.
The first-order path-conservative Osher scheme is recovered by replacing the limiter function \eqref{minmod} with the zero function. 
We note that the non-conservative upwind scheme is then recovered if we choose to evaluate the integrals of Eq.~\eqref{Osher} by using downwind-biased Riemann sums.

Empirically, the method is observed to be stable under the Courant--Friedrichs--Lewy (CFL) condition
\begin{equation}
	\text{Co} = \max_{\substack{0\leqslant i\leqslant N_x \\ 0\leqslant j\leqslant N_y}} \max \left\lbrace \varrho_{\mathbf{A}}(\mathbf{p}_{i,j}^n)\frac{\Delta t}{\Delta x},\, \varrho_{\mathbf{B}}(\mathbf{p}_{i,j}^n) \frac{\Delta t}{\Delta y} \right\rbrace \leqslant \frac1{2} \, ,
	\label{CFL2D}
\end{equation}
where $\text{Co}$ is the maximum Courant number in the $X$ and $Y$ directions. The spectral radii $\varrho_{\mathbf{A}} = \max |\mathbf{\Lambda}|$ and $\varrho_{\mathbf{B}}$ are deduced from the spectral decomposition \eqref{EigenA} of the system's matrices.
The stability of the scheme \eqref{SchemaExplCons} is also restricted by the spectral radius of the Jacobian matrix $\mathbf{S}'(\mathbf{p})$. For the examples presented hereinafter, a comparison of the stability limits implies that the scheme \eqref{SchemaExplCons} is stable under the classical CFL condition \eqref{CFL2D}. Hence, given a spatial discretization and a Courant number $\text{Co}$, the value of the time step $\Delta t$ is updated at each iteration according to Eq.~\eqref{CFL2D}.


\subsection{Validation}\label{sec:Validation}

The previous method is applied to a set of test cases with analytical solutions are detailed in \ref{app:Fund}. The first two tests correspond to the two-dimensional linear case where $\alpha = 1$, $\beta_2 = 0$, $\beta_3 = 0$, and $\rho$ and $\mu$ are taken  in Table~\ref{tab:Params}.
The third test is performed in a one-dimensional nonlinear case where $\alpha = 0$, $\beta_2 = 0$ (cubic nonlinearity only), and the other parameters are specified in Table~\ref{tab:Params}. The values in Table~\ref{tab:Params} are representative of pig brain matter \cite{destrade19, jiang15}. Unless stated otherwise, the Courant number is $\mathrm{Co} = 0.45$.

\begin{table}[h!]
	\centering
	
	{\renewcommand{\arraystretch}{1.2}
	\begin{tabular}{cccc}
		\toprule
		$\rho$ (kg/m\textsuperscript{3}) & $\mu$ (Pa) & $\beta_2$ & $\beta_3$ \\
		\midrule
		$1.04\times 10^3$ & $2.4\times 10^3$ & $-0.43$ & $1.41$ \\
		\bottomrule
	\end{tabular}}

	\caption{Mechanical parameters of a pig brain matter sample \cite{destrade19, jiang15}. 
	\label{tab:Params}}
	
\end{table}


\paragraph{Linear initial-value problem} The source term $\mathbf{S}_{i,j}^n$ is zero. The initial data $\mathbf{p}_{i,j}^0$ with wavelength $\lambda = 0.2$~m is obtained by computing the cell averages of
\begin{equation}
	\mathbf{p}^\circ(\mathbf{n}\cdot\bm{X}) = v^\circ\! \left(\frac{X + Y}{\lambda\sqrt{2}}\right) \begin{bmatrix}
	1 \\
	(1-\sqrt{3})/c\\
	(1-\sqrt{3})/c\\
	0 \\
	0 \\
	0
	\end{bmatrix}
	\qquad \text{with} \qquad
	v^\circ(\xi) = \left\lbrace \begin{aligned}
	&\cos^2(2\pi\xi) &&\text{if}\quad {-\tfrac34} < \xi < -\tfrac14,\\
	&1 &&\text{if}\quad \tfrac14 < \xi < \tfrac34,\\
	&0 &&\text{elsewhere}
	\end{aligned} \right.
	\label{CauchyLin}
\end{equation}
in m/s.
Thus, the initial data consists of a smooth sinusoidal bump and a rectangular bump, which propagate along the direction of angle $\varphi = \pi/4$ (see  \ref{app:Fund}).

\begin{figure}
	\centering
	
	\begin{minipage}{0.49\textwidth}
		\centering
		
		(a)
		\vspace{0.5em}
		
		\includegraphics{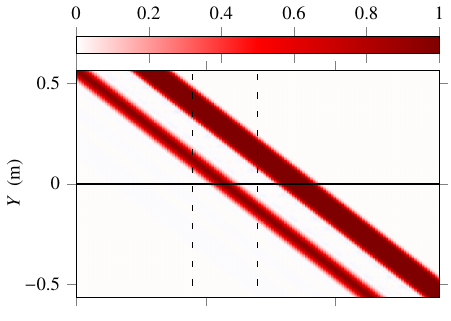}
		
		\vspace{-0.5em}
		
		\hspace{-0.2em}\includegraphics{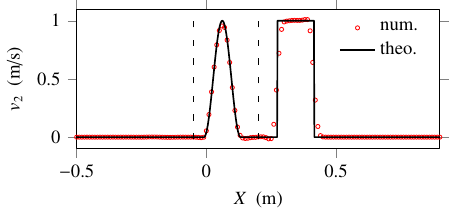}
	\end{minipage}
	\begin{minipage}{0.49\textwidth}
		\centering
		
		(b)\vspace{0.5em}
		
		\includegraphics{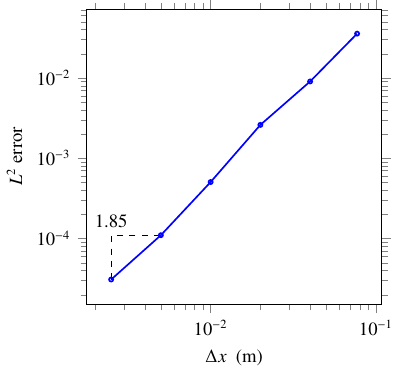}
	\end{minipage}
	
	\caption{Linear initial-value problem with initial data consisting of a smooth sinusoidal bump and a rectangular bump. (a) Top: map of the velocity $v_2$ in m/s obtained numerically. Bottom: cut along the line $Y=0$. (b) $L^2$ error measurements restricted to the line segment between dashed lines. \label{fig:CauchyLin}}
\end{figure}

In theory, the initial data is translated diagonally with constant speed $\tfrac{1 + \sqrt{3}}{2\sqrt{2}} c$. This is illustrated in Fig.~\ref{fig:CauchyLin}, which displays the numerical solution. It was obtained by iterating the time-stepping formula \eqref{SchemaExplCons} up to $t \approx 0.1$~s. Fig.~\ref{fig:CauchyLin}a displays the numerical solution obtained with $N_x = N_y = 200$. This figure shows that both parts of the wave are captured well. Fig.~\ref{fig:CauchyLin}b shows error measurements in $L^2$-norm performed along the line $Y = 0$ for ${-0.05}< X< 0.2$ (smooth bump), where $N_x = N_y$ varies from $25$ to $800$. The experimental order of accuracy is evidenced by the slope of the error curve. With the present method, second-order accuracy is obtained.

\paragraph{Linear non-homogeneous problem} The initial data $\mathbf{p}_{i,j}^0$ is zero. The only non-zero component of the body force $\bm f$ is $f_2 = 2\rho g$  (N/m$^3$). The corresponding non-zero component of the source term $\mathbf{S}$ is of the form $S_1 = g$ (see \ref{app:Matrices}). Here, we consider a sinusoidal point source $g(\bm{X},t) = a \delta(X)\delta(Y) s (t)$ with $s(t) = \sin(\omega t)$ for times $t> 0$. Computing the cell averages of $\mathbf{S}$, we have
\begin{equation}
	\mathbf{S}_{i,j}^n = a \frac{\delta_{i,i_0}}{\Delta x} \frac{\delta_{j,j_0}}{\Delta y}\, s(t_n) \begin{bmatrix}
		1 \\
		0\\
		0\\
		0 \\
		0 \\
		0
	\end{bmatrix}.
	\label{SourceLin}
\end{equation}
Indeed, the cell average of Dirac deltas $\delta(X)\delta(Y)$ produces Kronecker symbols $\delta_{i,i_0}\delta_{j,j_0}$ divided by the cell's surface area $\Delta x\Delta y$. Here, the source is localised at the origin, i.e. $x_{i_0} = 0$ and $y_{j_0} = 0$ when the numbers of cells $N_x$, $N_y$ are even integers. The source has amplitude $a = 0.2$~m$^3/$s$^2$ and angular frequency $\omega = 20\pi$~rad/s.

\begin{figure}[h!]
	\centering
	
	\begin{minipage}{0.49\textwidth}
		\centering
		
		(a)\vspace{0.2em}
		
		\includegraphics{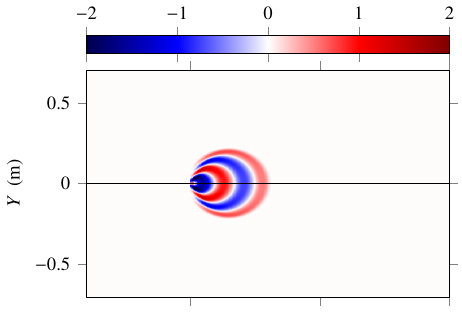}
		
		\vspace{-0.7em}
		
		\hspace{0.8em}~\includegraphics{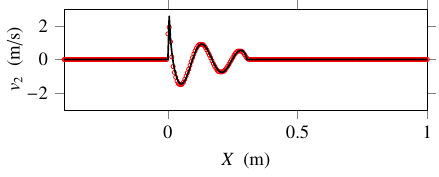}
	\end{minipage}
	\begin{minipage}{0.49\textwidth}
		\centering
		
		(b)\vspace{0.2em}
		
		\includegraphics{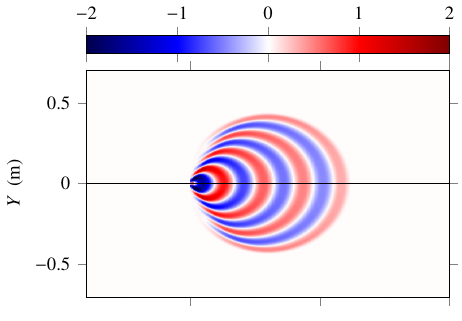}
		
		\vspace{-0.7em}
		
		\hspace{0.8em}~\includegraphics{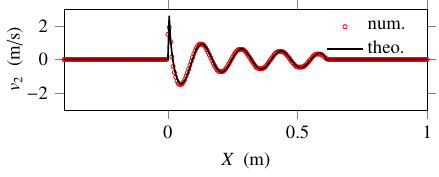}
	\end{minipage}
	
	\caption{Linear non-homogeneous problem with initial data consisting of a sinusoidal point source. Top: map of the velocity $v_2$ in m/s obtained numerically at $t\approx 0.2$~s (a) and $t\approx 0.4$~s (b). Bottom: comparison of the numerical and analytical solutions along the line $Y=0$. 
	\label{fig:SourceLin}}
\end{figure}

Figure~\ref{fig:SourceLin} represents the solution obtained numerically for $N_x = N_y = 400$, where the time-stepping formula \eqref{SchemaExplCons} was iterated up to $t \approx 0.4$~s. As time increases, a directional wave beam propagates along the $X$-direction. 

Comparisons between the numerical solution and the analytical solution of Eq.~\eqref{AnalyticConv} in Figs.~\ref{fig:SourceLin}a-\ref{fig:SourceLin}b show that the numerical method produces consistent results. Due to diffraction, the long-time velocity amplitude decreases as $X^{-1/2}$ with the distance of propagation $X$. The slight amplitude mismatch between numerical and analytical computations is due to the numerical diffusion of the MUSCL scheme. The small phase mismatch is caused by the explicit integration of the source. All these numerical artifacts vanish as the mesh is refined.

\paragraph{Nonlinear initial-value problem} This configuration is spatially one-dimensional, and the source term $\mathbf{S}_{i,j}^n$ is zero. The initial data $\mathbf{p}_{i,j}^0 = \mathbf{p}^\circ(x_i,y_j)$ with wavelength $\lambda = 0.4$~m reads
\begin{equation}
	\mathbf{p}^\circ(\bm{X}) = v^\circ(X/\lambda) \begin{bmatrix}
	1 \\
	0\\
	0\\
	0 \\
	0 \\
	0
	\end{bmatrix},
	\qquad \text{with} \qquad
	v^\circ(\xi) =
	\left\lbrace \begin{aligned}
	&1 &&\text{if}\quad {-\tfrac14} < \xi < \tfrac14,\\
	&\tfrac12 &&\text{elsewhere}
	\end{aligned} \right.
	\label{CauchyNL}
\end{equation}
in m/s.
The time-stepping formula \eqref{SchemaExplCons} is applied up to $t \approx 0.2$~s, with a Courant number $\mathrm{Co} = 0.95$. This value, larger than $0.5$, does not induce numerical instability due to the one-dimensional nature of the problem. The initial data \eqref{CauchyNL} consists of a rectangular bump, which is invariant along $Y$ (see  analytical developments in \ref{app:Fund}). 

\begin{figure}
	\centering
	
	\hspace{3em}
	\begin{minipage}{0.29\textwidth}
		\centering
		
		(a)
	\end{minipage}
	\begin{minipage}{0.29\textwidth}
		\centering
		
		(b)		
	\end{minipage}
	\begin{minipage}{0.29\textwidth}
		\centering
		
		(c)
	\end{minipage}
	
	\includegraphics{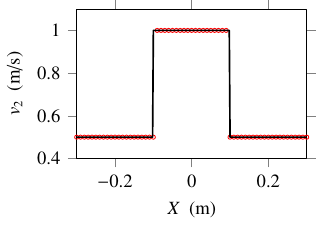}
	~~
	\includegraphics{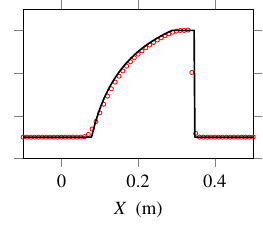}
	~~
	\includegraphics{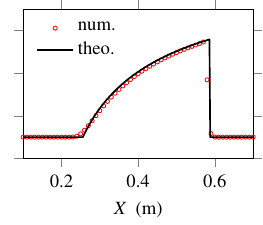}
	
	\caption{Nonlinear initial-value problem \eqref{CauchyNL} which is invariant along the $Y$-direction. Numerical and analytical solution displayed at $t=0$~s (a), $t=0.1$~s (b) and $t=0.2$~s (c). \label{fig:CauchyNL}}
\end{figure}

Similarly to Burgers' equation with rectangular data \cite{berjamin17}, the solution is made of a rarefaction and a shock, which interact after a certain amount of time. This is illustrated in Fig.~\ref{fig:CauchyNL}, where the solution obtained with $N_x = 200$ is displayed at three different times. The number of cells in the $Y$-direction includes twice the schemes stencil, i.e. $N_y = 5$. The position of the discontinuity in Fig.~\ref{fig:CauchyNL}c is obtained quasi-analytically, by numerical integration of the Rankine--Hugoniot condition. 
We  note that the method captures both the rarefaction and the shock wave, and that the latter is well-located.

\section{Harmonic generation in Gaussian beams}\label{sec:Harmonic}

The main goal of this section is to investigate harmonic generation numerically by solving boundary-value problems. This configuration is closely related to other studies in the literature, and it provides a natural way to control velocity amplitudes. Similarly to Destrade et al.~\cite{destrade19}, we consider two-dimensional motions with Gaussian sound beams.

The computational domain is reduced to $\bm{X} = (X,Y)$ in $[0 , 0.6] \times [ -0.6, 0.6] /\sqrt{2}$. Therefore, the mesh size is now deduced from $N_x = 0.6/\Delta x$ and $N_y = 1.2/(\Delta y \sqrt{2})$. The initial data $\mathbf{p}_{i,j}^0$ is zero, and the boundary data is specified at the domain's left boundary $X=0$. Numerically, the boundary condition is imposed by implementing an incoming wave condition (Sec. 7.3.2 of \cite{leveque02}). The domain's top, bottom and right boundaries have the same absorbing properties as in the previous section. Receivers are placed every $0.05$~m along the line $Y=0$  to record the signal in time.


\subsection{Linear case}

Here the nonlinearity coefficients $\beta_2$, $\beta_3$ are set to zero. The  boundary data is that of a pure anti-plane shear beam, with only non-zero component $v_2(0,Y,t) = a h(Y) s(t)$, where the spatial evolution is a Gaussian function $h(Y) = \exp({-(\frac\omega{c} Y)^2})$. In practice, the function $h(Y)$ is truncated at the distance $3\frac{c}\omega$ from the origin, where the Gaussian has sufficiently vanished. 
We take a causal sinusoidal signal $s(t) = \sin(\omega t)$ with amplitude $a = 0.7$~m/s and angular frequency $\omega = 20\pi$~rad/s. 

The solution is obtained numerically for $N_x = 250$ and $N_y = 500$, where the time-stepping formula \eqref{SchemaExplCons} was iterated up to $t \approx 0.6$~s. With the present grid, we have 63 points per wavelength at the fundamental frequency, and 10 points per wavelength at the sixth harmonic frequency. A snapshot of the final numerical solution is shown  in Figure~\ref{fig:HarmLin}a, and a video of the simulation is provided in the supplementary material.

\begin{figure}[h!]
	
	\centering
	
	\begin{minipage}{0.49\textwidth}
		\centering
		
		(a)\vspace{0.2em}
		
		\includegraphics{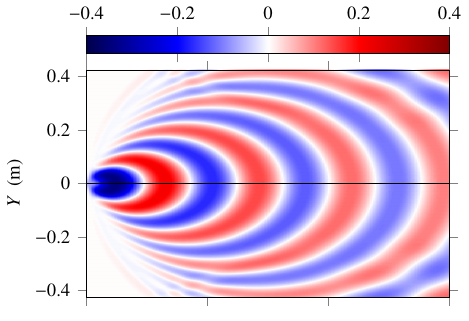}
		
		\vspace{-0.7em}
		
		\hspace{0.1em}~\includegraphics{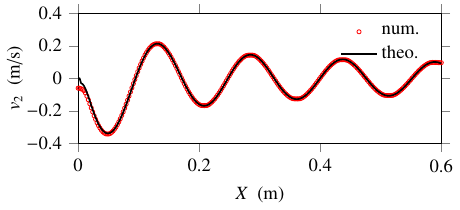}
		
	\end{minipage}
	\begin{minipage}{0.49\textwidth}
		\centering
		
		(b)\vspace{0.5em}
		
		\includegraphics{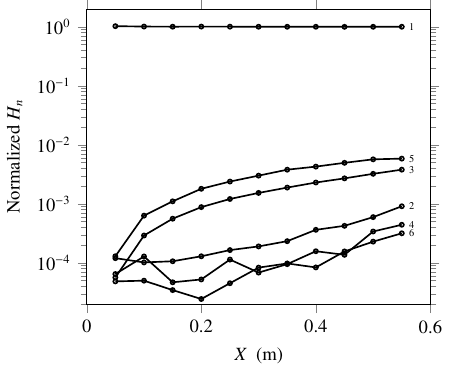}
		
	\end{minipage}
	
	\caption{Linear boundary-value problem. (a) Top: snapshot of the velocity $v_2$ at $t \approx 0.6$~s; Bottom: numerical and analytical solutions along the line $Y=0$. A video of the simulation is provided in the supplementary material. (b) Evolution of the normalized harmonic amplitudes with the propagation distance $X$ at $Y=0$ (labels: order $n$). \label{fig:HarmLin}}
\end{figure}

To estimate whether a given harmonic amplitude is significant or not, we measure the harmonic amplitudes along the beam axis. Figure~\ref{fig:HarmLin}b displays the evolution of the harmonic amplitudes with the propagation distance. The sine and cosine Fourier coefficients $a_n$, $b_n$ at the angular frequency $\omega$ were computed over the last period of signal by numerical integration (trapezoidal rule). Then, the harmonic amplitudes $H_n = \|({a_n}, {b_n})\|_2$ were divided by the theoretical harmonic amplitude $H_1$ of the first harmonic, see analytical solution \eqref{BeamGauss} in \ref{app:Fund}. At each receiver, a small amount of undesired harmonics{\,---\,}mainly odd ones{\,---\,}is spuriously generated by the numerical procedure. Therefore, in the nonlinear cases below, only harmonic amplitudes larger than those in Fig.~\ref{fig:HarmLin}b will be considered to be physically significant.


\subsection{Cubic nonlinearity only}

The nonlinearity coefficient $\beta_2$ is set to zero, while $\beta_3$ is taken from Table~\ref{tab:Params}. The other parameters are the same as in the linear case. Without the quadratic nonlinearity coefficient $\beta_2$, the system \eqref{SystWochner} governing displacement components decouples and $v_3$ remains equal to zero. 

\begin{figure}[h!]
	\centering
	
	\begin{minipage}{0.49\textwidth}
		\centering
		
		(a)\vspace{0em}
		
		\includegraphics{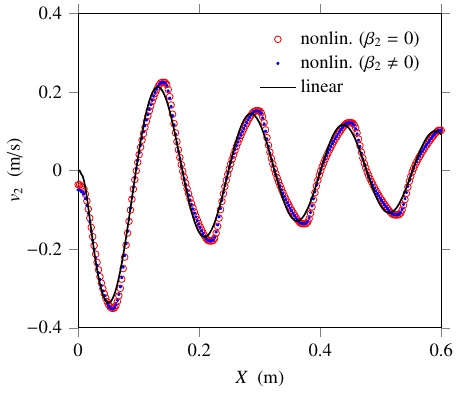}
		
	\end{minipage}
	\begin{minipage}{0.49\textwidth}
		\centering
		
		(b)\vspace{0em}
		
		\includegraphics{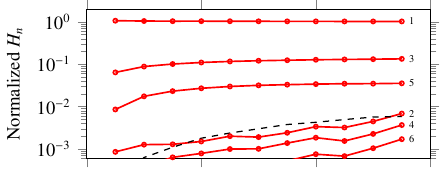}
		\vspace{-0.5em}
		
		\hspace{0.01em}
		\includegraphics{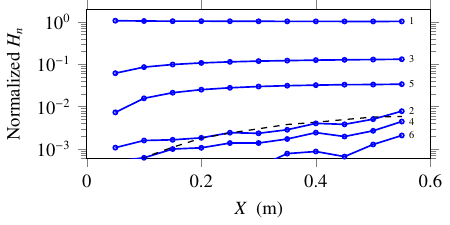}
		
	\end{minipage}
	
	\caption{Nonlinear boundary-value problem with purely cubic nonlinearity (red), or with quadratic and cubic nonlinearity (blue). (a) Velocity signal obtained numerically at $t\approx 0.6$~s; (b) Normalized harmonic amplitudes with respect to the propagation distance if $\beta_2=0$ (top) or if $\beta_2\neq 0$ (bottom). The dashed curve marks the fifth harmonic obtained in the linear case (Fig.~\ref{fig:HarmLin}b). \label{fig:Harm}}
\end{figure}

Figure~\ref{fig:Harm} illustrates the generation of odd harmonics with increasing propagation distances. In the farfield, numerical results show that harmonic generation slows down as waves propagate, due to the combined effects of nonlinearity and wave diffraction (diminution of wave amplitudes). In the absence of diffraction $(\alpha = 0)$, a shock would have formed at the distance $X_s \approx 0.08$~m \cite{zabo04} (see also \ref{app:Fund}). By making amplitudes decrease as $X^{-1/2}$, diffraction prevents wave breaking, and the solution keeps smooth during the simulation.

\subsection{Quadratic and cubic nonlinearity}

Both nonlinearity coefficients $\beta_2$, $\beta_3$ are taken from Table~\ref{tab:Params}. In the present configuration, the full system is solicited, and the results are included in Fig.~\ref{fig:Harm}. We note that the picture is very similar to the purely cubic case, up to the fact that the second harmonic is slightly more present. This observation confirms that such shear waves produce mainly odd harmonics, and it is consistent with the analysis of Destrade et al. \cite{destrade19} when the initial data is a pure anti-plane shear beam.


\section{Conclusion}\label{sec:Conclu}

We introduced a numerical method that solves the coupled nonlinear partial differential equations governing 2D directional shear waves in elastic solids. 
We proposed a change of variables leading to a hyperbolic system of first-order partial differential equations, where the time variable is the physical time $t$. The resulting system is in quasi-linear form, which suggests that specific numerical methods can be implemented. 
We showed that a MUSCL--Osher path-conservative Godunov-type method provides a second-order shock-capturing algorithm. The slope-limiting procedure prevents spurious oscillations from being produced around discontinuities. Numerical examples illustrate how the algorithm can be used to study the propagation of nonlinear shear-wave beams.

The method has great potential. It could be used to investigate how various polarizations of $v_2$ and $v_3$ lead to different harmonic generation features. As inferred by Destrade et al.~\cite{destrade19}, the second harmonic may be generated more substantially if the velocity field $v_3$ was not zero at the boundary.
According to their calculations, by enhancing the coupling between both components of the velocity field through the quadratic term, the second harmonic may reach magnitudes similar to the fifth harmonic.

The method could also be extended to other KZK-type equations which include dissipation \cite{pinton08, destrade11}. It could be extended to nonlinear viscoelastic and anisotropic materials \cite{meral10,lejeunes11,balbi18}, and to  slightly compressible materials. Moreover, the modelling of directional wave beams in pre-stressed solids{\,---\,}a.k.a. materials submitted to conditioning \cite{mechri16}{\,---\,}is an open problem \cite{destrade07}.

The implementation of such a method in the full three-dimensional case remains a challenge. The development of an efficient code using parallelization, higher-order methods and adaptive mesh refinement would be useful for applications \cite{castro06, reinarz20}.


\section*{Acknowledgments}

This work was supported by the Irish Research Council under project ID GOIPD/2019/328. The authors thank the Irish Centre for High-End Computing (ICHEC) for the provision of computational facilities and support.


\appendix

\section{Using the principal invariants}\label{app:Invar}

Let us express the principal invariants $\textit{I}_{\bm C}$, $\textit{II}_{\bm C}$, $\textit{III}_{\bm C}$ of the right Cauchy--Green tensor $\bm{C} = \bm{F}^\top\! \bm{F}$ in terms of the  invariants $I_k = \text{tr}\, \bm{E}^k$:
\begin{equation}
	\begin{aligned}
	\textit{I}_{\bm C} &= \text{tr}\,\bm{C} & \!\!\!\!\! & = 3 + 2 I_1 \\
	\textit{II}_{\bm C} &= \tfrac12 \big({\textit{I}_{\bm C}}^2 - \text{tr}\,\bm{C}^2\big) & \!\!\!\!\! &	= 3 + 4 I_1 + 2 {I_1}^2 - 2I_2 \\
	\textit{III}_{\bm C} &= \det \bm{C} & \!\!\!\!\! & = 1 + 2 I_1 + 2 {I_1}^2 - 2 I_2 + \tfrac43 {I_1}^3 - 4 I_1 I_2 + \tfrac83 I_3 ,
	\end{aligned}
	\label{IwithJ}
\end{equation}
or inversely,
\begin{equation}
	\begin{aligned}
	I_1 &= \tfrac12 (-3 + \textit{I}_{\bm C}) \\
	I_2 &= \tfrac14 (3 - 2 \textit{I}_{\bm C} - 2 \textit{II}_{\bm C} + {\textit{I}_{\bm C}}^2)  \\
	I_3 &= \tfrac18 (-3 + 3 \textit{I}_{\bm C} + 6 \textit{II}_{\bm C} - 3 {\textit{I}_{\bm C}}^2 + 3 \textit{III}_{\bm C} - 3 \textit{I}_{\bm C} \textit{II}_{\bm C} + {\textit{I}_{\bm C}}^3) \, .
	\end{aligned}
	\label{JwithI}
\end{equation}

The incompressibility constraint \eqref{Incomp} imposes $\textit{III}_{\bm C} \equiv 1$, which implies that one invariant $I_k$ depends on the two others.
Substituting the expressions \eqref{JwithI} in the strain energy function \eqref{WZabo} leads to a fourth-order Rivlin series \cite{destrade19}
\begin{equation}
	W = \sum_{i+j=1}^4 c_{ij} (\textit{I}_{\bm C} - 3)^i  (\textit{II}_{\bm C} - 3)^j
	\label{WZaboC}
\end{equation}
which coefficients are given in Table~\ref{tab:RivlinSeries}.
Using the chain rule along with the tensor derivatives of the principal invariants ${\partial \textit{I}_{\bm C}}/{\partial\bm C} = \bm{I}$ and ${\partial \textit{II}_{\bm C}}/{\partial\bm C} = \textit{I}_{\bm C} \bm{I} - \bm{C}$, the constitutive law is written as
\begin{equation}
	\bm{S} = -p\bm{C}^{-1}\! + 2\left(\frac{\partial W}{\partial \textit{I}_{\bm C}} + \textit{I}_{\bm C} \frac{\partial W}{\partial \textit{II}_{\bm C}}\right) \bm{I} - 2\frac{\partial W}{\partial \textit{II}_{\bm C}}\bm{C}
	\label{ConstSC}
\end{equation}
with the coefficients
\begin{equation}
	\begin{aligned}
	2\left(\frac{\partial W}{\partial \textit{I}_{\bm C}} + \textit{I}_{\bm C} \frac{\partial W}{\partial \textit{II}_{\bm C}}\right) &= -\mu + \tfrac14 A - \tfrac32 D + D \textit{I}_{\bm C} - \tfrac12 D {\textit{I}_{\bm C}}^2 - (\tfrac14 A-D) \textit{II}_{\bm C} \\
	- 2\frac{\partial W}{\partial \textit{II}_{\bm C}} \qquad\qquad &= \mu - \tfrac12 A + \tfrac32 D + (\tfrac14 A-D) \textit{I}_{\bm C} + \tfrac12 D {\textit{I}_{\bm C}}^2 - D \textit{II}_{\bm C} \, .
	\end{aligned}
	\label{ConstSCCoeffs}
\end{equation}
The above constitutive law is rewritten in terms of the Cauchy stress tensor $\bm{\sigma} = \bm{F}\bm{S} \bm{F}^\top\!$ in Eq.~\eqref{ConstSigB}.

\begin{table}
	
	\centering
	
	{\renewcommand{\arraystretch}{1.4}
		\begin{tabular}{cccccc}
			\toprule
			$i+j$ & $c_{ij}$ & & & & \\
			\midrule
			$1$   & $-\frac12\mu - \frac18 A$ & $\mu + \frac18 A$ & & & \\
			$2$   & $\tfrac14 D$ & $-\frac18 A - D$ & $\frac14\mu + \frac14 A + D$ & & \\
			$3$   & $0$ & $0$ & $-\tfrac14 D$ & $\tfrac1{24} A + \tfrac12 D$ & \\
			$4$   & $0$ & $0$ & $0$ & $0$ & $\tfrac1{16} D$ \\
			\bottomrule
	\end{tabular}}
	
	\caption{Coefficients $c_{ij}$ of the Rivlin series \eqref{WZaboC}, where the index $i$ increases{\,---\,}respectively, the index $j$ decreases{\,---\,}from the left ($i=0$) to the right ($j=0$). \label{tab:RivlinSeries}}
\end{table}

\section{System matrices}\label{app:Matrices}

The matrices of the first-order system in retarded time $\tilde t$ are specified below:
\begin{equation*}
	\mathbf{a} =
	\begin{bmatrix}
		1 & 0 & 0 & & & \\
		0 & 0 & 0 & & & \\
		0 & 0 & 0 & & & \\
		& & & 1 & 0 & 0 \\
		& & & 0 & 0 & 0 \\
		& & & 0 & 0 & 0 
	\end{bmatrix} , \qquad
	\mathbf{b} =
	\begin{bmatrix}
		0 & 0 & 0    & & & \\
		1 & 0 & 0    & & & \\
		0 & 0 & -\tfrac{\alpha^2}2 c & & & \\
		& & & 0 & 0 & 0 \\
		& & & 1 & 0 & 0 \\
		& & & 0 & 0 & -\tfrac{\alpha^2}2 c
	\end{bmatrix} , \qquad
	\mathbf{c}_\text{L} =
	\begin{bmatrix}
		0 & -1 & 0 & & & \\
		0 & 0 & -1 & & & \\
		0 & 1 & 0  & & & \\
		& & & 0 & -1 & 0 \\
		& & & 0 & 0 & -1 \\
		& & & 0 & 1 & 0 
	\end{bmatrix} ,
\end{equation*}
\begin{equation}
	\mathbf{c}_\text{NL}(\mathbf{q}) =
	\frac{\beta_2}{2 c}\!
	\begin{bmatrix}
		0 & 0 & 0 & & & \\
		0 & 0 & 0 & & & \\
		0 & 0 & 0 & -\vartheta & 0 & v_3 \\
		& & & 0 & 0 & 0 \\
		& & & 0 & 0 & 0 \\
		-\vartheta & 0 & v_3 & 2\varepsilon & 0 & -2v_2
	\end{bmatrix}
	- \frac{\beta_3}{3 c^3}\!
	\begin{bmatrix}
		0 & 0 & 0 & & & \\
		0 & 0 & 0 & & & \\
		3{v_2}^2 + {v_3}^2 & 0 & 0 & 2v_2v_3 & 0 & 0 \\
		& & & 0 & 0 & 0 \\
		& & & 0 & 0 & 0 \\
		2v_2v_3 & 0 & 0 & {v_2}^2 + 3{v_3}^2 & 0 & 0 
	\end{bmatrix} .
\end{equation}
The source term has components $\mathbf{s} = \frac12 c/\mu\, (0,0, f_2,0,0, f_3)^\top\!$.
The matrices $\mathbf{A}(\mathbf{p}) = [A_{ij}]$, $\mathbf{B}(\mathbf{p}) = [B_{ij}]$ and the vector $\mathbf{S}(\mathbf{p}) = [S_i]$ of the quasi-linear system of balance laws \eqref{QLSystemCons} in real time $t$ are deduced from the above arrays. Non-zero coefficients are detailed below:
\begingroup
\begin{align}
	A_{11} &=  \tfrac{1}{c\Delta}\big(1 + \beta_2\varepsilon - \tfrac13 \beta_3\tfrac{{v_2}^2 + 3 {v_3}^2}{c^2}\big) & A_{14} &= \tfrac{1}{c\Delta}\big(\tfrac12 \beta_2\vartheta + \tfrac23\beta_3 \tfrac{v_2v_3}{c^2} \big) = A_{41} \nonumber\\
	A_{21} &= -1 = A_{54} & A_{44} &= \tfrac{1}{c\Delta}\big(1 - \tfrac13 \beta_3\tfrac{3{v_2}^2 + {v_3}^2}{c^2}\big)
\nonumber\\
	B_{11} &= \beta_2 \tfrac{v_3}{c^2 \Delta} \big( \tfrac14{\beta_2}\vartheta + \tfrac13\beta_3\tfrac{v_2v_3}{c^2} \big) & B_{31} &= -1 = B_{64} \nonumber\\
	B_{13} &= -\tfrac12\alpha^2 \tfrac{1}{\Delta} \big(1 + \beta_2 \varepsilon - \tfrac13 \beta_3 \tfrac{{v_2}^2 + 3{v_3}^2}{c^2}\big) &  B_{41} &= \tfrac12 \beta_2 \tfrac{v_3}{c^2 \Delta} \big(1 - \tfrac13\beta_3 \tfrac{3{v_2}^2 + {v_3}^2}{c^2}\big) \nonumber\\
	B_{14} &= \tfrac12 \beta_2 \tfrac{v_3}{c^2 \Delta} \big( 1 + \beta_2\varepsilon - \tfrac13 \beta_3 \tfrac{5 {v_2}^2 + 3 {v_3}^2}{c^2} \big) - \tfrac12 {\beta_2}^2 \tfrac{v_2}{c^2 \Delta} \vartheta & B_{44} &= -\beta_2\tfrac{v_2}{c^2\Delta} \big(1 - \tfrac13\beta_3 \tfrac{3 {v_2}^2 + 2 {v_3}^2}{c^2}\big) + \tfrac14 {\beta_2}^2 \tfrac{v_3}{c^2 \Delta} \vartheta \\
	B_{16} &= -\alpha^2 \tfrac{c^2}{\beta_2 v_3} B_{11} = B_{43} & B_{46} &= -\alpha^2 \tfrac{c^2}{\beta_2 v_3} B_{41} \nonumber
\end{align}
\begin{equation*}
	\begin{aligned}
		S_1 &= \tfrac{1}{2\mu \Delta} \big(f_2 + \tfrac12 \beta_2 (2 f_2\varepsilon + f_3\vartheta) + \tfrac13\beta_3\tfrac{2v_2 (f_2 v_2 +  f_3 v_3) - 3 f_2 ({v_2}^2 + {v_3}^2)}{c^2}\big) \\
		S_4 &= \tfrac{1}{2\mu \Delta} \big(f_3 + \tfrac12 \beta_2 f_2 \vartheta + \tfrac13\beta_3\tfrac{2v_3 (f_2 v_2 +  f_3 v_3) - 3 f_3 ({v_2}^2 + {v_3}^2)}{c^2}\big)
	\end{aligned}
\end{equation*}
\endgroup
and the determinant $\Delta = \det \mathbf{c}'(\mathbf{q})$ is expressed in Eq.~\eqref{DeterminantC}.


\section{Some analytical solutions}\label{app:Fund}


\subsection{Linear waves}

Here the coefficients $\beta_2$, $\beta_3$ are equal to zero. Therefore, the motion \eqref{SystWochner} is governed by the PDE $(u_{, t} + c u_{, 1})_{, t} - \tfrac{1}2 \alpha^2 c^2 u_{, 22} = g$, where the right-hand side $g = \tfrac{1}2 f /\rho$ represents a density of force per unit mass (in m/s\textsuperscript{2}).
Introducing the partial derivatives $v = u_{, t}$, $\gamma = u_{, 1}$ and $\varepsilon = u_{, 2}$ in a similar manner to Sec.~\ref{sec:FirstOrder}, the first-order system of conservation laws
\begin{equation}
	\begin{bmatrix}
		v\\
		\gamma\\
		\varepsilon
	\end{bmatrix}_{, t} + \begin{bmatrix}
		c & 0 & 0 \\
		-1 & 0 & 0 \\
		0 & 0 & 0
	\end{bmatrix}
	\begin{bmatrix}
		v\\
		\gamma\\
		\varepsilon
	\end{bmatrix}_{, 1} + \begin{bmatrix}
		0 & 0 & -\tfrac{1}2 \alpha^2 c^2 \\
		0 & 0 & 0 \\
		-1 & 0 & 0
	\end{bmatrix}
	\begin{bmatrix}
		v\\
		\gamma\\
		\varepsilon
	\end{bmatrix}_{, 2}
	= \begin{bmatrix}
		g\\
		0\\
		0
	\end{bmatrix}
	\label{LinearSyst}
\end{equation}
is obtained. It is of the form $\mathbf{p}_{, t} + \mathbf{A} \mathbf{p}_{,1} + \mathbf{B} \mathbf{p}_{,2}  = \mathbf{S}$ with $\mathbf{p} = (v, \gamma, \varepsilon)^\top\!$. In what follows, we present some particular solutions to homogeneous initial- and boundary-value problems ($g \equiv 0$), and to the non-homogenenous problem $g \not \equiv 0$ with zero initial conditions.

\paragraph{Initial value problems}

In this paragraph, we assume that $g \equiv 0$. We consider initial value problems of the form $\mathbf{p}(\bm{X},0) = \mathbf{p}^\circ(\bm{X})$, where $\bm{X} = (X,Y)$ is the vector of spatial coordinates. Particular solutions can be obtained in the eigenspaces of the matrix $\mathbf{M} = n_1\mathbf{A} + n_2\mathbf{B}$ where $\mathbf{n} = (n_1, n_2)$ is a unit vector.
The spectrum of $\mathbf{M}$ reads $\lbrace 0, \lambda_\pm\rbrace$ with
\begin{equation}
\lambda_\pm = \frac{c}2 \left( \cos\varphi \pm \sqrt{\cos^2\!\varphi + 2\alpha^2\sin^2\!\varphi}\right) ,
\end{equation}
where the angle $\varphi$ satisfies $\mathbf{n} =  (\cos\varphi, \sin\varphi)$. If $\cos\varphi$ and $\alpha$ both equal zero, then $\mathbf{M}$ is not diagonalizable. Otherwise, several eigenspaces can be identified.
\begin{itemize}
	\item the kernel of $\mathbf{M}$ has dimension one or two, and any vector in this eigenspace is of the form $\mathbf{p} = (0, \gamma, \varepsilon)^\top\!$. The system yields $\gamma_{,t} = 0$ and $\varepsilon_{,t} = 0$. Therefore, if the initial data is of the form $\mathbf{p}^\circ\! = (0, \gamma^\circ\! , \varepsilon^\circ)^\top\!$, then the solution of the initial-value problem is $\mathbf{p}(\bm{X}, t) = \mathbf{p}^\circ(\bm{X})$. The solution is stationary.
	
	\item
	we consider vectors $\mathbf{p} = \varepsilon\, (\tfrac{c^2}2 \alpha^2 \sin^2\!\varphi/\lambda_\mp, \cos\varphi, \sin\varphi)^\top\!$ in the eigenspace of $\mathbf{M}$ corresponding to a nonzero eigenvalue $\lambda_\pm$. As $\alpha \sin \varphi\to 0$, this eigenvector becomes $\mathbf{p} \to \varepsilon\, (-\lambda_\pm, \cos\varphi, \sin\varphi)^\top\!$ and the eigenvalue becomes $\lambda_\pm \to c\cos\varphi$, where $\pm$ corresponds to the sign of $\cos\varphi$. The system yields $\varepsilon_{,1}\sin\varphi = \varepsilon_{,2} \cos\varphi$, which implies that $\varepsilon$ is a function of $\mathbf{n\cdot} \bm{X}$ and $t$. Also, the system provides $\varepsilon_{,t}\sin\varphi + \lambda_\pm \varepsilon_{,2} = 0$ and $\varepsilon_{,t} \cos\varphi + \lambda_\pm \varepsilon_{,1} = 0$. Thus, if $\mathbf{p}^\circ$ is of the present form and can be expressed as a function of $\mathbf{n\cdot} \bm{X}$, then the solution to the initial value problem is $\mathbf{p}(\bm{X},t) = \mathbf{p}^\circ(\mathbf{n\cdot} \bm{X} - \lambda_\pm t)$. The solution is a plane wave propagating along the $\mathbf{n}$-direction with the speed $\lambda_\pm$.
\end{itemize}
Duhamel's principle and the Green's function could be used to derive more general solutions, of a similar kind to those in the next paragraphs.

\paragraph{Non-homogeneous problems}

We expand Green's function for the second-order scalar form of \eqref{LinearSyst}, where zero initial conditions for $u$, $u_{,t}$ are considered. For the computation of the two-dimensional Green's function $u = G$, we define the source term as $g(\bm{X},t) = a\delta(X)\delta(Y)\delta(t)$ where $\delta$ is the Dirac delta. The amplitude $a \neq 0$ is expressed in m\textsuperscript{3}/s. Fourier transformation of the PDE leads to 
\begin{equation}
	\big({-\omega}^2 + c \kappa_X \omega + \tfrac{1}2 \alpha^2 c^2 {\kappa_Y}^2\big)\, \mathcal{F} [\hat G] = a \, ,
	\label{LinearFourier}
\end{equation}
where $\omega$ denotes the angular frequency and $\bm{\kappa} = (\kappa_X,\kappa_Y)$ is the vector of spatial frequencies in the $\bm{X}$-direction. Here, the hat symbol denotes Fourier transformation $\int (\cdot)\, \text{e}^{\text i \omega t} \text d t$ in the time domain, and the operator $\mathcal{F}$ denotes the spatial Fourier transform $\iint (\cdot)\, \text{e}^{-\text i \bm{\kappa}\cdot \bm{X}} \text d\bm{X}$.
Solutions in Fourier domain may be obtained if the polynomial factor in \eqref{LinearFourier} is nonzero, which is assumed from now on. Using the definition of the time-domain Fourier transform, we have
\begin{equation}
	\mathcal{F} [G] = \frac{a}{2\pi} \int_{\mathbb R}\! \frac{\text{e}^{-\text i \omega t}  \text d \omega}{(\omega_+-\omega)(\omega-\omega_-)} \, , \qquad
	\omega_\pm = \frac{c}2 \left( \kappa_X \pm \sqrt{{\kappa_X}^2 + 2 \alpha^2 {\kappa_Y}^2} \right) .
	\label{LinearFourierInv}
\end{equation}
The integral in Eq.~\eqref{LinearFourierInv} is evaluated as part of a contour integral in the complex $\omega$-plane (a half circle in the lower half of the complex plane which includes $\omega_\pm$). Taking the limit as the radius of the contour increases to infinity, the residue theorem yields
\begin{equation}
	\mathcal{F} [G] = a\, \frac{\text{e}^{-\text i \omega_- t} - \text{e}^{-\text i \omega_+ t}}{\text i \,(\omega_+ - \omega_-)} \qquad\text{for}\qquad t>0 \, ,
	\label{LinearFourierRes}
\end{equation}
and the inverse spatial Fourier transform $G = (2\pi)^{-2}\! \iint \mathcal{F} [G] \, \text{e}^{\text i \bm{\kappa}\cdot\bm{X}} \text d \bm{\kappa}$
provides an integral representation of the fundamental solution $G(X,Y,t)$.

If $\alpha = 0$, the evaluation of the fundamental solution is rather straightforward, and we have
\begin{equation}
	G(X,Y,t) = \frac{a}{c} \delta(Y)  \operatorname{H}(X) \operatorname{H}(ct-X) \, ,
	\label{FundAdv}
\end{equation}
where $\operatorname{H}$ denotes the Heaviside function.
If $\alpha \neq 0$, the evaluation of the fundamental solution is more involved.
To do so, let us introduce stretched polar coordinates $\bm{\kappa} = k\, \big(\! \cos\theta, \tfrac1{\alpha\sqrt2} \sin\theta \big)$ such that $\text d\bm{\kappa} = \tfrac1{\alpha\sqrt2} k \, \text d k\, \text d \theta$. The fundamental solution is rewritten as
\begin{equation}
	G(X,Y,t) = \frac{a}{\pi c \alpha\sqrt{2}} \int_0^\infty \sin (\tfrac12 kct) \left[ \frac{1}{2\pi} \int_0^{2\pi} \text{e}^{\text i (\bm{\kappa} \cdot \bm{X} - \frac12 \kappa_X c t)} \text d \theta \right] \! \text d k \, .
	\label{IntegralRep}
\end{equation}
Then, by identifying various integral representations \cite{toscano13}, we find
\begin{equation}
	G(X,0,t) = \frac{a}{\pi c \alpha \sqrt{2}} \, \frac{\operatorname{H} (X) \operatorname{H} (ct - X)}{\sqrt{ X (c t - X)}}
	\label{Bessel3}
\end{equation}
for $Y = 0$.
This solution becomes singular as $\alpha \to 0$, which is coherent with Eq.~\eqref{FundAdv}.

Now let us consider a general point source $g(\bm{X},t) = a \delta(X)\delta(Y) s(t)$, where $s$ is a causal dimensionless signal. For sake of dimensional homogeneity, the coefficient $a$ is expressed in m\textsuperscript{3}/s\textsuperscript{2} here. Forward Fourier transformation of the PDE leads to the identity $\mathcal{F} [\hat u] = \mathcal{F} [\hat G] \, \hat s$ in Fourier domain. By virtue of the convolution theorem, backward Fourier transformation leads to $u = G *_t s$, where $*_t$ denotes convolution in time. Differentiation in time then leads to $v = G *_t s'$, where $s'$ is the causal derivative of $s$. In the case $\alpha= 0$, the classical expression
\begin{equation}
	v(X,Y,t) = \frac{a}{c} \delta(Y) \operatorname{H}(X)\, s(t-X/c)
	\label{AnalyticConvAdv}
\end{equation}
is deduced from the expression \eqref{FundAdv} of $G$. In the case $\alpha\neq 0$, the velocity is represented by
\begin{equation}
	v(X,0,t) 
	= \frac{a}{c \alpha \sqrt{2\pi c}} \frac{\operatorname{H} (X)}{\sqrt{X}}\, D^{1/2} s(t-X/c) \, ,
	\label{AnalyticConv}
\end{equation}
where $D^{1/2}$ denotes the Caputo derivative of order 1/2, defined by
\begin{equation}
	\begin{aligned}
		D^{1/2} s(\tilde t) = \frac{1}{\sqrt{\pi}}\int_{0}^{\tilde t} (\tilde t - \tau)^{-1/2} s'(\tau)\, \text d\tau = \frac{2}{\sqrt{\pi}}\int_{0}^{\sqrt{\tilde t}} \!\! s'(\tilde t - \vartheta^2)\, \text d\vartheta \, .
	\end{aligned}
	\label{AnalyticCaputo}
\end{equation}
In practice, the change of variable $\vartheta = (\tilde t - \tau)^{1/2}$ and then numerical integration are used to evaluate the fractional derivative.

Now, let us consider a causal periodic point source, for instance such that $s(t) = \sin(\omega t)$ for positive times. In the case $\alpha = 0$, Eq.~\eqref{AnalyticConvAdv} leads to $v(X,Y,t) = \frac{a}{c} \delta(Y) \sin(\omega \tilde t)$ for positive $X$ and positive $\tilde t = t-X/c$. In the case $\alpha \neq 0$, we observe that the solution \eqref{AnalyticConv} is asymptotically periodic as the time goes to infinity. In fact, as described in \cite{vigue19}, we may express the velocity field as
\begin{equation}
	v(X,0,t) \simeq
	\frac{a}{c \alpha} \frac{\operatorname{H} (X)}{\sqrt{X}}\, \sqrt{\frac{\omega}{2\pi c}}\, \sin\!\big(\omega (t-X/c) + \tfrac{\pi}{4}\big)
	\label{AnalyticFourier}
\end{equation}
for large times $t\to {+\infty}$. In the next paragraph, an extension of \eqref{AnalyticConv}--\eqref{AnalyticFourier} to the whole domain is introduced.

\paragraph{Boundary value problems}

In this paragraph, we assume that $g \equiv 0$.
We consider a boundary value problem of the form $v(0,Y,t) = a \delta(Y) s(t)$ with $a$ in $\text{m}^2/\text{s}$. The causal signal $s$ is dimensionless. Let us transform back to retarded time $\tilde t = t-X/c$. The boundary value problem reads $v_{,1\tilde t} = \frac12 c\alpha^2 v_{,22}$ with $v(0,Y,\tilde t) = a \delta(Y) s(\tilde t)$ for positive times $\tilde t$. Fourier transformation in space and retarded time gives
\begin{equation}
	\big(\kappa_X \tilde\omega + \tfrac{1}2 \alpha^2 {\kappa_Y}^2 c\big)\, \mathcal{F} [\hat v] = 0 \, .
	\label{LinearFourierBC}
\end{equation}
Non-trivial solutions are obtained if the dispersion relation $\kappa_X \tilde\omega = -\tfrac12 \alpha^2 {\kappa_Y}^2 c$ is satisfied.
Partial Fourier transformation of the PDE with respect to $Y$ and $\tilde t$ leads to a first-order boundary value problem for $\mathcal{F}_Y [\hat v]$ in terms of $X$, where $\mathcal{F}_Y$ denotes the Fourier operator $\int (\cdot)\, \text{e}^{-\text i \kappa_Y Y} \text dY$. After partial integration with respect to $X$, we end up with $\mathcal{F}_Y [\hat v] = a \hat s(\tilde \omega) \text{e}^{\text i\kappa_X X}$ for positive $X$, where $\kappa_X$ is deduced from the dispersion relation. Inverse Fourier transformation in $Y$ and $\tilde t$ then provides the integral representation
$v = a\, (2\pi)^{-2}\! \iint \hat s(\tilde \omega) \text{e}^{\text i (\bm{\kappa}\cdot \bm{X} - \tilde \omega \tilde t)} \text d\kappa_y\text d\tilde \omega$
of the solution.

If $\alpha = 0$, the evaluation of the solution is rather straightforward, and the classical expression
\begin{equation}
	v(X,Y,\tilde t) = a\, \delta(Y) \operatorname{H}(X)\, s(\tilde t)
	\label{FundAdvBV}
\end{equation}
is recovered. Up to a multiplicative coefficient, this expression is the same as Eq.~\eqref{AnalyticConvAdv}. If $\alpha \neq 0$, then integration along $\kappa_Y$ amounts to the computation of generalized Gaussian integrals. Indeed, completing the squares in the exponentials leads to integrals of the form $\int \text e^{\pm\text i k^2}\text d k$, which are common in quantum field theory (Appendix~A of \cite{zee10}). After integration w.r.t. $\kappa_Y$, we find
\begin{equation}
	v(X,Y,\tilde t) = \frac{a}{\alpha\sqrt{2\pi c}} \frac{\operatorname{H}(X)}{\sqrt{X}} \left(\frac{1}{2\pi}\int_{\mathbb R} (-\text i \tilde \omega)^{1/2} \hat{s}(\tilde\omega) \, \text e^{-\text i \tilde \omega \left(\tilde t - \frac12 \frac{Y^2}{\alpha^2 c X}\right)}\text d\tilde\omega \right) = \frac{a}{\alpha\sqrt{2\pi c}} \frac{\operatorname{H}(X)}{\sqrt{X}}\, D^{1/2} s\!\left(\tilde t - \tfrac12 \tfrac{Y^2}{\alpha^2 c X}\right) ,
	\label{FundBV}
\end{equation}
where the coefficient $(-\text i \tilde\omega)^{1/2}$ is the symbol of the fractional derivative $D^{1/2}$ in time domain. Up to a factor $c$, the solution \eqref{FundBV} of the boundary-value problem coincides with the solution of the non-homogeneous problem\footnote{Duhamel's principle provides a proof of this property over the whole domain (for $Y\neq 0$ in particular). To do so, replace $a$ by $\frac{a}{c} \delta(\xi)$ in the expression of the boundary data and in Eq.~\eqref{FundBV}. Then, consider the velocity $\int_{0}^X v(X-\xi,Y,t)\, \text d\xi$ with $v(X,Y,t)$ deduced from \eqref{FundBV} to solve the non-homogeneous problem.} \eqref{AnalyticConv} along the line $Y=0$. In the case of periodic forcing, the long-time solution is obtained by following \cite{vigue19}, in a similar manner to the non-homogeneous problem \eqref{AnalyticConv}--\eqref{AnalyticFourier}.

The fundamental solution \eqref{FundBV} can be used to solve more general problems with boundary data of the form $v(0,Y,t) = a h(Y) s(t)$. Using the convolution theorem for the coordinate $Y$, we find $v = h *_Y v_\delta$ where $v_\delta$ is the expression in Eq.~\eqref{FundBV} obtained for $h = \delta$. Alternatively, we may write $\mathcal{F}_Y[v] = \mathcal{F}_Y[h]\, \mathcal{F}_Y[v_\delta]$ in Fourier domain. In particular, if $h \equiv 1$, then evaluation of the convolution product gives $v = a \operatorname{H}(X)\, h(Y) s(\tilde t)$. This result is obvious given that the present problem is invariant along the $Y$-coordinate.  Now, consider a monochromatic Gaussian beam where $h(Y) = \exp({-(\frac\omega{c} Y)^2})$ and $s(\tilde t) = \sin(\omega\tilde t)$ for all $\tilde t$. The Fourier transform $\mathcal{F}_Y[h]$ is a Gaussian integral, and $\mathcal{F}_Y[v_\delta]$ follows from $\mathcal{F}_Y [\hat v_\delta] = a \hat s(\tilde \omega) \text{e}^{\text i\kappa_X X}$. Taking the inverse Fourier transform of $\mathcal{F}_Y[v] = \mathcal{F}_Y[h]\, \mathcal{F}_Y[v_\delta]$ leads again to generalized Gaussian integrals \cite{zee10}. Finally, we find
\begin{equation}
	v(X,Y,\tilde t) = \frac{a \operatorname{H}(X)}{\sqrt[4]{1+x^2}} \exp\! \left({-\tfrac1{1+x^2}}(\tfrac\omega{c} Y)^2\right) \sin\! \left(\omega \tilde t - \tfrac{x}{1 + x^2} (\tfrac\omega{c} Y)^2 + \arctan\!\big(\tfrac{\sqrt{1+x^2} - 1}{x}\big)\right)
	\label{BeamGauss}
\end{equation}
with $x = 2\alpha^2\frac\omega{c} X$. The near-field and far-field regions are characterized by $\frac{\omega}{c} X \ll 1$ and $\frac{\omega}{c} X \gg 1$, respectively. In the near-field range, the diffraction-free expression $v \simeq a \operatorname{H}(X)\, h(Y) s(\tilde t)$ is recovered. In the far-field range, we recover the long-time solution deduced from the point source \eqref{FundBV}.

\subsection{Nonlinear waves}

Consider the homogeneous system \eqref{QLSystemCons} with $\mathbf{S} = \mathbf{0}$. Here, both the diffraction and the quadratic nonlinearity are neglected ($\alpha = 0$ and $\beta_2 = 0$), but the cubic nonlinearity has coefficient $\beta_3>0$. Alternatively, assume that the configuration is invariant along the transverse $Y$-axis. In retarded time, these assumptions lead to a set of two coupled Burgers-like equations satisfied by $v_2$, $v_3$ \cite{zabo04, wochner08}. In physical time, a similar process leads to the set of equations
\begin{equation}
	\begin{aligned}
	\bar v_{2, 1} + \tfrac1{c} \bar v_{2,t} &= \tfrac1{3c} \beta_3 \big(\bar v_2 ({\bar v_2}^2 + {{\bar v_3}}^2)\big)_{, t} \, ,\\
	\bar v_{3, 1} + \tfrac1{c} \bar v_{3,t} &= \tfrac1{3c} \beta_3 \big(\bar v_3({{\bar v_2}}^2 + {{\bar v_3}}^2)\big)_{, t} \, ,
	\end{aligned}
	\label{BurgersSyst}
\end{equation}
where $\bar v_i = v_i/c$.
The Riemann invariants $\bar w_1 = {\bar v_2}^2 + {\bar v_3}^2$ and $\bar w_2 = {\bar v_3}/{\bar v_2}$ of \eqref{BurgersSyst} satisfy the scalar transport equations $\bar w_{i,1} + \sigma_i \bar w_{i,t} = 0$, where the characteristic slownesses $\sigma_i$ are given by $\sigma_1 = \frac1{c} (1 - \beta_3\bar w_1)$ and $\sigma_2 = \frac1{c} (1 - \frac13 \beta_3\bar w_1)$.
The characteristic field with slowness $\sigma_1$ is genuinely nonlinear everywhere except at the origin, whereas the characteristic field with slowness $\sigma_2$ is linearly degenerate. One notes that $\bar w_1$ represents the squared modulus in the complex $\bar v_2$-$\bar v_3$ plane, while $\bar w_2$ is related to the argument.
In what follows, several analytical methods are briefly introduced. Interested readers are referred to the literature for complements \cite{godlewski96, lax73}.

\paragraph{Initial value problems}
The initial value problem $\mathbf{p}(\bm{X},0) = \mathbf{p}^\circ(\bm{X})$ can be solved analytically in terms of the Riemann invariants $\bar w_1$ and $\bar w_2$ of \eqref{BurgersSyst}, which satisfy the scalar transport equations $\bar w_{i,t} + \bar w_{i,1}/\sigma_i = 0$. Solutions to smooth initial value problems can be expressed in implicit form up to the breaking time by applying the method of characteristics. In particular, if the invariant $\bar w_1$ is constant in space and time, then both velocity components of \eqref{BurgersSyst} are transported at constant speed $1/\sigma_2${\,---\,}that is to say, waves propagate linearly at the same speed.
Similarly, if we assume that $\bar w_2$ is a constant, then the velocity components of \eqref{BurgersSyst} are advected non-linearly at the speed $1/\sigma_1$ with $\bar w_1 = (1+{\bar w_2}^2){\bar v_2}^2$ (or equivalently, $\bar w_1 = (1+{\bar w_2}^{-2}){\bar v_3}^2$). The system \eqref{BurgersSyst} decouples, and it can be rewritten in conservation form as
\begin{equation}
	\begin{aligned}
	&\bar v_{2, t} + F_{+1}  (\bar v_{2})_{ ,1} = 0 \\
	&\bar v_{3, t} + F_{-1} (\bar v_{3})_{ ,1} = 0 
	\end{aligned}
	\qquad\text{where}\qquad
	F_k (\bar v) = c\bar v\, \frac{\operatorname{artanh}\! \left(\bar v\sqrt{(1+{\bar w_2}^{2k}) \beta_3}\right)}{\bar v\sqrt{(1+{\bar w_2}^{2k}) \beta_3}} 
	\label{Burgers}
\end{equation}
and where $\operatorname{artanh}$ is the inverse hyperbolic tangent function. Thus, we note that the linear advection equation is recovered at small amplitudes. Solving the Riemann problem of \eqref{Burgers} for shock and rarefaction waves requires particular care, since the $\operatorname{artanh}$ function is neither convex nor concave (see e.g. Ref.~\cite{berjamin17a} and references therein). To avoid complications, the example considered in this document involves data located on the same side of the inflection point.

\paragraph{Boundary value problems}
The same method applies for the boundary value problem $\mathbf{p}(0,Y,t) = \mathbf{p}^\circ(t)$. The Riemann invariants $\bar w_1$ and $\bar w_2$ of \eqref{BurgersSyst} satisfy $\bar w_{i,1} + \sigma_i\bar w_{i,t} = 0$. Solutions to smooth boundary value problems can be expressed in implicit form up to the shock distance. If the only nonzero component at the boundary is $v_2(0,Y,t) = a \sin (\omega t)$, then the shock distance reads $X_s = {c^3}/({\beta_3 \omega a^2})$ \cite{lax73, zabo04}.

\end{document}